\documentclass[fleqn,10pt]{wlscirep}  
\usepackage{xr}  
\usepackage[utf8]{inputenc}
\usepackage{units}
\usepackage{mathtools}  
\usepackage{amsmath}
\usepackage{amssymb}
\usepackage{commath}  
\usepackage{subcaption}
\usepackage{comment}
\usepackage[T1]{fontenc}
\usepackage{lineno}

\begin{document}

\title{Bio-inspired decision making in robot swarms under biases}

\author[1,*]{Raina Zakir}
\author[2]{Timoteo Carletti}
\author[1]{Marco Dorigo}
\author[3,4,5,*]{Andreagiovanni Reina}
\affil[1]{IRIDIA, Université Libre de Bruxelles, Brussels, Belgium}
\affil[2]{Department of Mathematics and Namur Institute for Complex Systems, naXys, University of Namur, Namur, Belgium}
\affil[3]{Centre for the Advanced Study of Collective Behaviour, Universit\"{a}t Konstanz, Konstanz, Germany}
\affil[4]{Department of Computer and Information Science, Universit\"{a}t Konstanz, Konstanz, Germany}
\affil[5]{Department of Collective Behaviour, Max Planck Institute of Animal Behavior, Konstanz, Germany}
\affil[*]{raina.zakir@ulb.be; andreagiovanni.reina@uni-konstanz.de}



\begin{abstract} 
To operate autonomously, minimal robot swarms must make timely and reliable collective decisions despite noisy individual sensing and severe constraints on communication, computation, and memory. Achieving this capability could expand their use in applications such as healthcare, disaster response, and environmental monitoring. Here, we study how such swarms can rapidly and reliably reach consensus on the best among $n$ discrete options by comparing two canonical mechanisms of opinion dynamics---direct-switch and cross-inhibition---simple yet effective rules for collective information processing observed in biological systems across scales, from neural populations to insect colonies. We generalise existing mean-field models by incorporating asocial biases that influence opinion dynamics. 
While swarms using direct-switch reliably select the best option in the absence of asocial dynamics, their performance deteriorates when such biases are introduced, often leading to decision deadlocks. In contrast, bio-inspired cross-inhibition enables faster, more cohesive, robust, and scalable decisions across a wide range of biased conditions. Our findings provide theoretical and practical insights into the coordination of minimal swarms, with implications for a broad class of decentralised decision-making systems across biology and engineering.

\end{abstract}
\flushbottom
\maketitle
%
%
\thispagestyle{empty}



\section*{Introduction}
 
Robot swarms are decentralised multi-robot systems in which large numbers of robots coordinate through local interactions with one another and the environment, without centralised supervision~\cite{book:Hamann:2018}. Their distributed organization avoids single points of failure and allows the system to remain operational even when individual robots fail, making swarm systems particularly suitable for deployment in constrained or hazardous environments~\cite{zapata2020}, with potential applications ranging from targeted drug delivery~\cite{hauert2014nano} to the assessment of fire-affected infrastructure for containment planning~\cite{Tzoumas2024SI}. To achieve autonomy in such decentralised systems, enabling the swarm to make collective decisions is an essential requirement. 

In many real-world scenarios, robot swarms must select the best among alternatives that differ in quality or cost---a task formalised as the best-of-$n$ problem~\cite{Valentini2017Review}. 
Individual robots often operate with limited sensing and processing capabilities, and their observations may therefore be noisy or incomplete. Through repeated interactions and information exchange, however, swarms can integrate these uncertain signals and reach reliable collective decisions. Similar decision processes occur in biological collectives, which achieve robust, decentralised consensus by following simple local rules. For instance, ant and honeybee colonies routinely choose the best location to build their nest, with individual insects acting as noisy information-processing units \cite{seeley|buhrman:2001}. Analogously, populations of neurons integrate noisy signals to guide behavioural responses \cite{usher2001}. 
Despite vast differences in scale and substrate, these systems rely on similar mechanisms for information integration \cite{marshall2009,PIRRONE202266,Reina:scirep:2018}. Such mechanisms have also been adopted to design robot swarm behaviours for best-of-$n$ decision-making \cite{Reina:SwInt:2021,march2024honeybee}.
However, when deployed in real robot swarms, these mechanisms must remain robust to disturbances and asocial influences, a challenge that remains insufficiently addressed~\cite{Hunt2020}.

Across many natural collectives and in our robot swarms, individuals communicate their current opinion at a rate proportional to the perceived quality of the option they support~\cite{parker|zhang:2009,marshall|bogacz|dornhaus|planque|kovacs|franks:2009}. As a result, support for higher-quality options accumulates more rapidly, increasing the likelihood of consensus on the best option.
However, the resulting collective dynamics depend not only on how often opinions are exchanged, but also on how individuals integrate information from peers, particularly in the presence of asocial influences. In this paper, we compare two mechanisms for integrating peer information, commonly observed in natural systems and implemented in swarm robotics: direct-switch and cross-inhibition (Figure~\ref{fig:robot-fsm}A-B).

The direct-switch mechanism (Figure~\ref{fig:robot-fsm}A) periodically updates a robot's opinion by copying the one of a randomly chosen neighbour. 
This mechanism is commonly implemented through the weighted voter model, which is mathematically simple to analyse and computationally inexpensive to run and therefore widely adopted for solving the best-of-$n$ problem~\cite{valentini|hamann|dorigo:2014,valentini|ferrante|hamann|dorigo:2015,talamaliSciRob2021,shan2021discrete,palina2019}. 
The cross-inhibition mechanism \cite{pais|etal:2013,reina|valentini|fernandezoto|dorigo|trianni:2015,Reina:PRE:2017} (Figure~\ref{fig:robot-fsm}B), inspired by the house-hunting behaviour of honeybees~\cite{seeley|etal:2012}, is another prominent collective decision-making mechanism with simplicity comparable to direct-switch. 
Similar inhibitory interactions recur across several levels of biological organisation, enabling decisive winner-take-all selection among competing alternatives. For example, regulatory networks governing the cell cycle G2/M transition rely on inhibitory feedback loops that generate sharp, switch-like transitions between alternative states~\cite{Cardelli2012,Cardelli2017}. Likewise, at the neural level, decision dynamics are often described by the leaky competing accumulator model, whose mutual-inhibition architecture is formally analogous to cross-inhibition in collective decisions~\cite{marshall2009,PIRRONE202266,usher2001}. 
In swarm robotics, cross-inhibition is implemented such that when a robot receives a conflicting opinion from a neighbour (i.e., an opinion different from its current one), it does not copy it directly but instead becomes uncommitted (or undecided). Only uncommitted robots adopt the opinions of their neighbours. 

Both cross-inhibition and direct-switch enable units of the swarm---robots, animals, or neurons---to integrate social information and reach consensus on the best alternative, even when only a slight majority correctly senses the environment \cite{parker|zhang:2009,marshall2009,Reina:PRE:2017,valentini|ferrante|hamann|dorigo:2015}.
To meaningfully compare these two mechanisms, however, it is essential to study them under more realistic conditions, moving beyond idealised noise-free models, as real-world swarms are inevitably subject to bias, noise, and external interference.

In our framework, asocial dynamics occur when robots update their opinion based on information that does not reflect the state of their peers. This may occur when robots combine social interactions with independent environmental exploration, or when they receive biased, distorted, or misleading information.
While some forms of such asocial dynamics can be beneficial in controlled amounts~\cite{Prasetyo2019, zakir_miscommunication_nodate}, they more often impair collective performance by hindering consensus formation, prolonging indecision~\cite{Mobilia_2015,reina_cross-inhibition_2023,Khaluf2017noise}, or leading to suboptimal decisions~\cite{AntZakDorRei2024:aamas,giulia2021}. As illustrated in Figure~\ref{fig:bias}A-C, we investigate three representative sources of asocial dynamics: (i)~biased information spread by asocial robots (stubborn robots or zealots) that resist conforming to social pressure~\cite{reina_cross-inhibition_2023}, (ii)~spontaneous opinion changes driven by self-sourced environmental information~\cite{zakir2022}, and (iii)~corrupted communication due to message interception and alteration~\cite{zakir_miscommunication_nodate}.

Asocial dynamics can be categorised as unbiased---affecting all options equally---or biased towards certain options (Figure~\ref{fig:bias}D-E). Previous work has studied the direct-switch and cross-inhibition mechanisms in several specific cases that our general model can describe in a unified compact form, including unbiased social noise~\cite{Golman2014,Jhawar2020,Liu2025}, equally sized subgroups of zealots that can hinder or sometimes facilitate group consensus~\cite{reina_cross-inhibition_2023,Canciani2019,Prasetyo2019,JuliaKlein2024,Khalil_2018}, and miscommunication among robots~\cite{zakir_miscommunication_nodate,damore2025}. In particular, synergistic bias---where asocial dynamics favour the superior option---has mainly been studied in the form of self-sourcing (i.e., spontaneous discovery), where agents are more likely to independently encounter and commit to the best option~\cite{zakir2022,talamaliSciRob2021,Reina:PRE:2017,marshall2009,pais|etal:2013}. While it is often reasonable to assume that better options are closer or more abundant, and thus easier to find (e.g., highest net payoff when travel or search costs matter), this assumption does not always hold true. Antagonistic bias---where asocial dynamics favour the inferior option---can also frequently arise in both natural and artificial systems.
For instance, house-hunting insects~\cite{franks2008AnimBeh} and robot swarms~\cite{reina|valentini|fernandezoto|dorigo|trianni:2015,leaf2024} may face environments where the optimal choice is also the hardest to discover.
Human groups~\cite{Centola2018}, fish flocks~\cite{couzin2011}, and ant colonies~\cite{Rajendran2022} can include minorities of stubborn individuals with strong biases that sometimes steer the consensus~\cite{Marvel2012}. Artificial systems may also be exposed to cyberattacks~\cite{Hunt2020}, including communication tampering~\cite{zakir_miscommunication_nodate,damore2025} or stubborn robots supporting inferior options~\cite{giulia2021,AntZakDorRei2024:aamas,Canciani2019}. In previous studies, we analysed several of these asocial influences separately, showing that zealots, miscommunication, and self-sourcing can each critically alter collective dynamics and, in some cases, even improve decision accuracy under cross-inhibition~\cite{reina_cross-inhibition_2023,zakir_miscommunication_nodate,zakir2022}. Here, we unify these seemingly distinct forms of asocial dynamics within a single mathematical formulation, enabling a systematic comparison of decision-making mechanisms across the full spectrum from synergistic to antagonistic bias (see Supplementary Text ST1 for an extended discussion).

In this work, we investigate how asocial dynamics influence collective decision-making in robot swarms. 
We develop deterministic ordinary differential equation (ODE) models to describe the swarm population dynamics at multiple levels of abstraction. 
To capture stochastic and finite-size effects, we complement this analysis with a microscopic chemical-reaction-network model simulated using the Gillespie algorithm~\cite{gillespie2013perspective}, as well as robot swarm simulations and experiments. Our multiscale analysis shows that stubborn agents, corrupted communication, and independent information discovery can be represented by mathematically equivalent terms. This unified formulation (Figure~\ref{fig:bias}A-E) enables 
systematic comparison of decision-making mechanisms across the full spectrum from synergistic to antagonistic bias, in both binary and multi-option settings. Beyond robotics, the bio-inspired nature of these mechanisms offers insights into decision-making trade-offs in natural systems, including insect colonies and neural assemblies.
These results provide a general framework for analysing how different sources of asocial influence shape collective decision-making in distributed systems.

\section*{Results}

In the proposed framework, asocial dynamics can arise from any of the three factors illustrated in Figure~\ref{fig:bias}A-E. Rather than reading a message from a cooperative (social) peer, a robot may instead (i)~interact with a zealot---an agent persistently favouring one option~\cite{reina_cross-inhibition_2023,AntZakDorRei2024:aamas} (Fig.\,\ref{fig:bias}A); (ii)~self-source information from the environment, independently of social cues~\cite{zakir2022} (Fig.\,\ref{fig:bias}B); or (iii)~receive a corrupted message due to a man-in-the-middle (MITM) attack that intercepts and alters communication~\cite{zakir_miscommunication_nodate} (Fig.\,\ref{fig:bias}C). We generalise these distinct forms of asocial dynamics by modelling them as variants of a single underlying mechanism that occurs with probability~$\eta$ at each opinion update (see Methods~\nameref{sec:equivalence}). 
Under this formulation, each opinion update is based on representative social information with probability~$1-\eta$ (right branch in Fig.~\ref{fig:bias}F), and on asocial information with probability~$\eta$ (left branch in Fig.~\ref{fig:bias}F). Asocial dynamics can be biased towards any of the $n$ options, with probabilities $\eta_i$ for $i \in \{1,\dots,n\}$, where $\eta_i$ specifies how often an asocial update favours option $i$ and $\sum_{i=1}^{n}\eta_i = 1$. For the best-of-$2$ case, illustrated in Fig.~\ref{fig:bias}F, this reduces to $\eta_a$ and $\eta_b=1-\eta_a$.

\begin{figure}[!ht]
  \centering
 \includegraphics[width=\textwidth,trim={0cm 0cm 0cm 0cm},clip]{images/bias2qf.png}
      \caption{(A-C) We consider three types of asocial dynamics impacting the collective decision-making in robot swarms. (A)~Asocial agents: stubborn robots (shown on the left) broadcast their opinion and influence others (top row) but ignore incoming messages (bottom row). (B)~Asocial discovery: robots occasionally update their opinion by independently self-sourcing information from the environment (e.g., by sensing the ground colour in the collective perception scenario~\cite{zakir2022}) rather than through social interactions. (C)~Communication corruption: under man-in-the-middle (MITM) attacks, the content of exchanged messages is altered (the speech balloon and received opinion's colours are inconsistent), resulting in the spread of misinformation. (D-E)~Environmental evidence can be biased towards either option: (D)~synergistic bias favours the best option, while (E)~antagonistic bias favours the inferior one. Although illustrated here for asocial discovery, synergistic or antagonistic bias can also be present in the other asocial dynamics, i.e., different numbers of asocial agents supporting each option or biased message corruption. (F)~Unified schematic of how a robot supporting option A updates its opinion. With probability $1-\eta$, the robot updates from representative social information sampled from neighbours, with support for A or B weighted by the relative proportions of robots in states $A$ and $B$. With probability $\eta$, the update is instead driven by asocial information arising from asocial agents, communication corruption, or asocial discovery. In this case, the information is biased towards A or B with probabilities $\eta_a$ and $\eta_b$, respectively. Information supporting A leaves the robot's opinion unchanged. Instead, if the information supports B, and thus conflicts with the robot's current opinion, the transition depends on the decision mechanism: with direct-switch, the robot switches to $B$; with cross-inhibition,  socially sourced conflicting information always leads to the uncommitted state ($U$), whereas asocially sourced conflicting information either leads to $U$ (type-1) or to a direct switch to $B$ (type-2).}
  \label{fig:bias}
\end{figure}

\subsection{Collective perception with a minimalistic robot swarm}

\begin{figure}[!ht]

\centering

\includegraphics[width=1\textwidth,trim={0cm 0cm 0cm 0cm},clip]{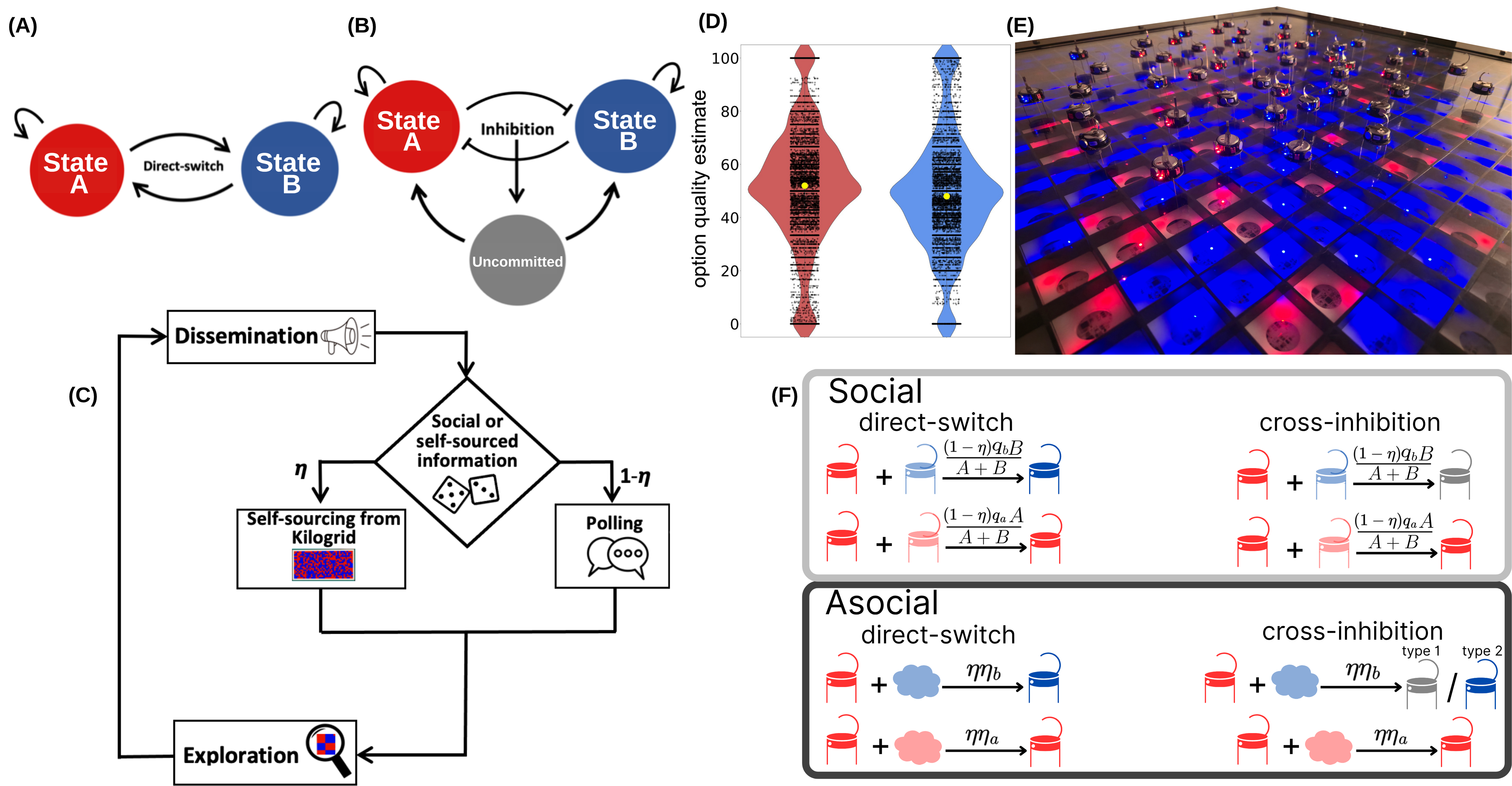}
\caption{Individual-level decision-making mechanisms. 
(A-B) The basic algorithmic representation of the two mechanisms: direct-switch (A) and cross-inhibition (B). They describe how an individual updates its opinion in response to social information and capture core interaction rules underlying both our robot-swarm algorithms and related mechanisms in natural systems. In direct-switch (A), a robot immediately adopts the received opinion. In cross-inhibition (B), a committed robot that receives an opposing opinion becomes uncommitted (i.e., it is inhibited). Only uncommitted robots can be recruited to either option. 
(C)~Finite state machine (FSM) describing robot behaviour, adapted from Valentini et al.~\cite{valentini|hamann|dorigo:2014} and extended here to include asocial dynamics corresponding to self-sourcing environmental information~\cite{zakir2022}. Rectangles denote FSM states and arrows denote transitions. After dissemination, robots update their opinion using either social information (by polling a neighbour) or self-sourced environmental information, before returning to the exploration state. 
(D)~Strip plots overlaid on violin plots showing 7\,500 individual quality estimates made by simulated robots for options A (red) and B (blue). Yellow markers indicate the ground-truth proportion of red and blue tiles in the Kilogrid arena. 
(E)~Experiment with 50 Kilobots~\cite{rubenstein|ahler|hoff|cabrera|nagpal:2014} operating on the Kilogrid arena~\cite{ValAntTra-etal2018:si} composed of red and blue tiles. 
(F)~Equivalent chemical-reaction-network representation of the collective decision-making process. Transitions occur at rates indicated above each arrow and are triggered either by encounters with social robots (lighter-coloured robots in the top box) or by asocial events (coloured clouds in the bottom box).}  \label{fig:robot-fsm} 

\end{figure}

We employ the investigated decentralised decision-making mechanisms to enable efficient collective perception in a minimalistic robot swarm.
As illustrated in Figure~\ref{fig:robot-fsm}A-B, robots can be committed to option A (state $A$) or option B (state $B$); under the cross-inhibition mechanism (Figure~\ref{fig:robot-fsm}B), robots may also be uncommitted (state $U$). Committed robots broadcast their opinion to nearby peers, while uncommitted robots remain silent. In both direct-switch and cross-inhibition mechanisms, robots periodically read a message from a randomly selected neighbour and update their opinion accordingly (Figure~\ref{fig:bias}F). 
For cross-inhibition, we consider two variants of the response to asocial events: after receiving asocial information (e.g., through self-sourcing), robots either become uncommitted (type-1) or switch directly to the new opinion (type-2).

Figures~\ref{fig:robot-fsm}A–B show basic algorithmic representations of the individual-level decision mechanisms, whose simplicity enables insights that generalise across domains, from robotics to natural collectives. Implementing these mechanisms in physical robots, however, requires adaptations to account for real-world constraints. A critical aspect is quality estimation. Determining the quality of an option may take time, either because robots must visit spatial locations~\cite{talamaliSciRob2021,valentini|ferrante|hamann|dorigo:2015} or repeatedly sample distributed environmental features~\cite{Hamann:ANTS:2026,chinIROS2023ImperfectCPerception}. 
Following earlier approaches~\cite{zakir2022,valentini|hamann|dorigo:2014,valentini|ferrante|hamann|dorigo:2015}, we therefore extend the basic algorithms by introducing two alternating phases: exploration and dissemination (Figure~\ref{fig:robot-fsm}C and Supplementary Text ST2). During exploration, robots estimate the quality of their currently supported option; during dissemination, they broadcast their opinion to neighbours.

Our robots are minimalistic, with extremely limited computation, memory, and communication capabilities. Like house-hunting insects~\cite{seeley|etal:2012,marshall2009}, each robot stores and communicates information about only one option. The robot's estimates can be extremely noisy, e.g., compare the true quality (yellow point) with the robot measurements (black points) in Figure~\ref{fig:robot-fsm}D. Despite these individual-level limitations, by disseminating an opinion for a time proportional to the estimated quality, the swarm collectively amplifies small differences in population opinions and reliably makes consensus decisions on the best option.

In our collective perception task~\cite{Hamann:ANTS:2026}, robots must choose between two environmental features, modelled as different floor tile colours (red for option A and blue for option B---see Figure~\ref{fig:robot-fsm}E). As they move, robots collect noisy measurements of local quality. Occasionally, a robot disregards social input and instead self-sources information from the environment by switching its commitment to the option associated with the tile it locally senses. Therefore, the proportion of tile colours determines the self-sourcing bias.

We consider two collective perception scenarios. The first, commonly studied in swarm robotics~\cite{valentini|brambilla|hamann|dorigo:2016,zakir2022,ebert2018multi,Hamann:ANTS:2026}, requires the swarm to determine the predominant ground colour. In this case, the option qualities $q_a$ and $q_b$ correspond to the proportion of tiles of each colour. This setup leads to synergistic asocial dynamics, because the probability of self-sourcing an option equals its quality ($\eta_a = q_a$ and $\eta_b = q_b$); therefore, if red tiles are more abundant, then both $q_a > q_b$ and $\eta_a > \eta_b$. In the second scenario, we decouple option quality from tile distribution in order to study both synergistic and antagonistic biases. Here, the self-sourcing bias $\eta_a$ is determined by the tile colour proportions, whereas $q_a$ and $q_b$ are independent quantities that reflect the intrinsic quality of each option as estimated by the robots during exploration. This decoupling enables us to study antagonistic bias ($\eta_a < 0.5$), where the inferior option is more frequently self-sourced because it is more abundant in the environment (e.g., more blue tiles despite red being higher quality).

\subsection{Multiscale modelling: from macroscopic population dynamics to robot swarm experiments}\label{sec:multiscale}

We model the collective decision-making process at multiple levels of abstraction by combining mechanistic descriptions, deterministic mean-field models, and stochastic simulations. The basic interaction rules are illustrated in Figures~\ref{fig:bias}F and~\ref{fig:robot-fsm}F. Building on these rules, we develop mathematical models by using tools from statistical mechanics~\cite{vankampen:1992} and dynamical systems to describe how the fractions of the population committed to options A and B evolve over time.

Equations~\eqref{eq:ode:vm:generic} and~\eqref{eq:ode:ci:generic:1}-\eqref{eq:ode:ci:generic:2} define the basic ODE models of direct-switch and cross-inhibition, respectively. These models capture the essential dynamics while abstracting from implementation-specific details (as in Figure~\ref{fig:robot-fsm}A-B). We also formulate robot-specific ODE models, presented in the Supplementary Material (Equations SE1 and SE2), which are tailored to the robotic implementation by including the alternating phases of environmental exploration (for quality estimation) and dissemination (for opinion broadcasting) shown in Figure~\ref{fig:robot-fsm}C. 
To capture stochasticity arising from finite swarm size, we also model the system as chemical-like reaction networks, which define the system's master equations (Supplementary Text ST7). This stochastic description complements the deterministic ODE models, which assume an infinitely large population.

\begin{figure}[!ht]
  \centering\includegraphics[height=0.66\textwidth,trim={0cm 0cm 0cm 0cm},clip]{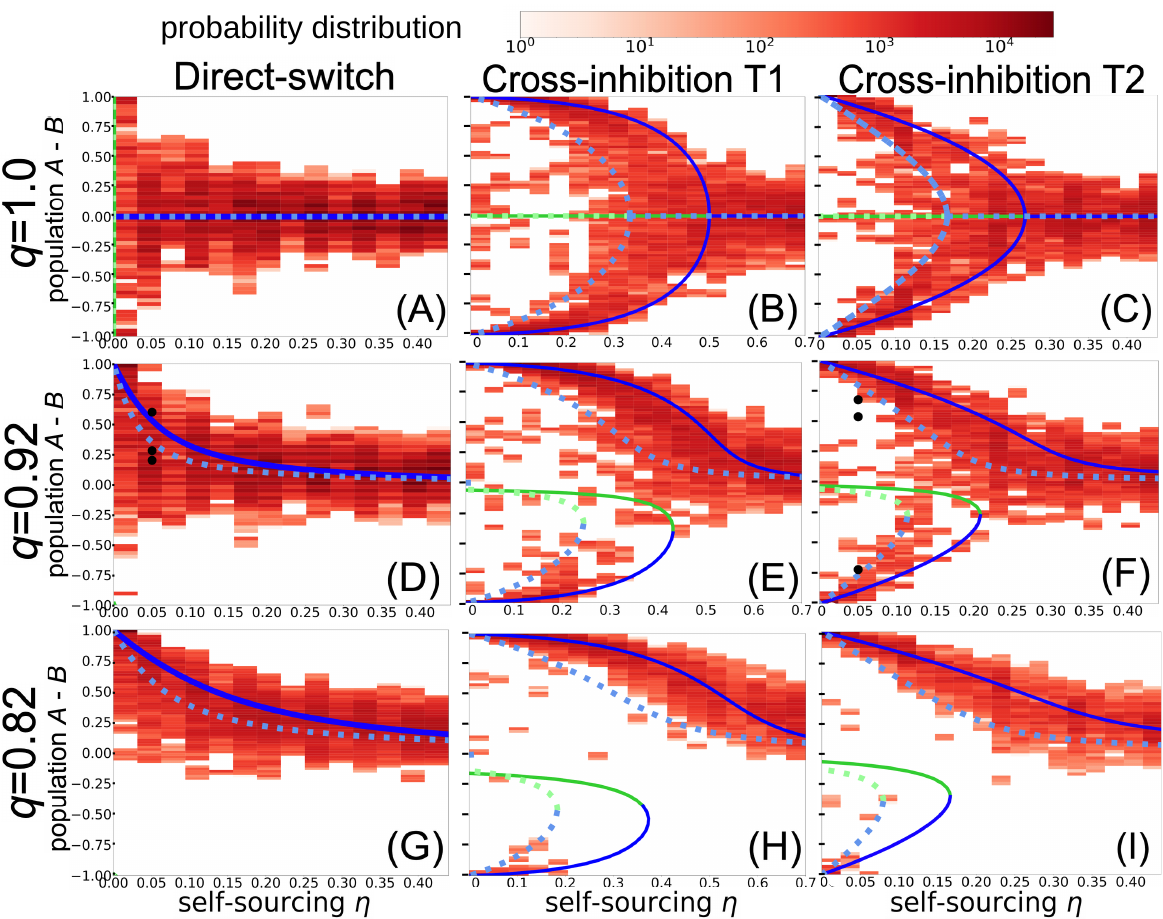}
  \caption{Results from swarm robotics simulations, physical robot experiments, and ODE models. Difference between the fractions of robots committed to options A and B ($A-B$, y-axis) as a function of asocial dynamics $\eta$ (x-axis) for quality ratios $q\in\{1,0.92,0.82\}$ (rows), and for direct-switch and cross-inhibition models (columns), in the collective-perception scenario with synergistic bias $\eta_a=q_a$, where both quality and self-sourcing favour the most abundant tile colour~\cite{zakir2022}. Cross-inhibition types T1 and T2 correspond to the cases shown in Figure~\ref{fig:bias}F. Red 2D histograms are computed from the last $1\,000$ time steps (i.e., \unit[6]{s}) of each of 50 robot-simulation runs for each $\eta$, and report the difference between the fractions of robots committed to option $A$ and $B$, i.e., (number robots for $A$ - number robots for $B$) / $N$. Black dots in D and F show the final fractions from selected real-robot experiments. Superimposed lines show ODE fixed points (green: unstable; blue: stable); dashed lines correspond to the basic models (Equations~\eqref{eq:ode:vm:generic} and~\eqref{eq:ode:ci:generic:1}-\eqref{eq:ode:ci:generic:2}), and solid lines to the robot-specific models (Equations SE1 and SE2). Basic ODE models are initialised with agents equally distributed between states $A$ and $B$, while the robot-specific ODE models are initialised with $t_d$/($t_e$+$t_d$) agents in dissemination states (split between $A_D$ and $B_D$), and the rest equally in $A_E$ and $B_E$. In the robot-specific models and simulations, $t_e=3\,000$, $t_u=1\,000$, and $t_d=1\,300$ (see Methods). Although different, the basic and robot-specific models share qualitatively the same dynamics, which are accurate predictions of the robot simulations (especially for the latter models). When option qualities are identical (A) or similar (D), direct-switch rarely reaches a stable majority, even under weak asocial dynamics. By contrast, cross-inhibition yields cohesive decisions across a broad range of $\eta$, regardless of the quality difference.}
  
  \label{fig:density_synergetic}
  
\end{figure}

Mean-field models describe the macroscopic dynamics of the system by capturing the time evolution of the fractions of robots committed to option A or B. We analyse these models via equilibrium and stability analyses of the ODEs, where the stable equilibria indicate the long-term collective behaviour of the swarm for $T \rightarrow \infty$. Figure~\ref{fig:density_synergetic} presents bifurcation diagrams showing how the stable and unstable equilibria vary with the frequency of asocial dynamics ($\eta$). Each diagram shows both basic and robot-specific ODE equilibria (dotted and solid lines, respectively), overlaid with the final state of robot simulations, shown as a red 2D histogram. While the basic ODEs capture the robot swarm behaviour only qualitatively, the robot-specific models show close quantitative agreement with the simulations.

We further validated these predictions with 24 experiments using 50 physical robots in a subset of representative conditions (Figure~\ref{fig:robot-fsm}E and Supplementary Videos\footnote{Main video available at: \url{https://youtu.be/CSirFypT9tY} and all videos available at \url{https://iridia.ulb.ac.be/supp/IridiaSupp2026-001/}.}), with the final fractions of robots committed to either option shown as overlaid black markers in Figures~\ref{fig:density_synergetic} and~\ref{fig:density_antagonistic}. Despite the additional sources of variability inherent to embodied robots, the experimental results closely match the trends predicted by the models and observed in simulation. Small deviations observed near full consensus are likely due to embodiment effects, such as robots becoming trapped near the arena boundaries because of their limited locomotion capabilities.

To further validate our models, we compare the stationary probability distribution (SPD) obtained from the chemical reaction networks with the final state of robot simulations and ODE equilibria (Supplementary Figure SF3 and SF9). The SPD, computed using the Gillespie algorithm~\cite{gillespie2013perspective}, captures the long-run probability distribution over all possible proportions of the swarm committed to options A and B. We find good agreement among the SPD, the robot simulation results, and the ODE model equilibria, demonstrating the consistency and robustness of our multiscale modelling framework.

\subsection{Fast, cohesive, accurate, and scalable collective decisions} 

Our analysis focuses on four key aspects of collective decision-making: group cohesion, decision accuracy, decision speed, and scalability, all of which are important performance metrics in both robotic and natural systems~\cite{franks|dornhaus|fitzsimmons|stevens:2003,CouzinEtAl2005,valentini|ferrante|hamann|dorigo:2015}.
While decentralised systems have been extensively studied in terms of speed--accuracy trade-offs~\cite{talamali2019improving,shan2021discrete, valentini|ferrante|hamann|dorigo:2015}, speed-cohesion trade-offs and scalability have received comparatively little attention~\cite{franks|etal:2013,Reina:PRE:2024}. We address this gap by systematically comparing the two mechanisms across all four metrics.

\subsubsection{Cohesion and symmetry breaking} 

A critical metric of collective decision-making is cohesion, defined as the proportion of agents committed to the majority opinion.
Stability analysis of the ODE models (Equations~\eqref{eq:ode:vm:generic} and~\eqref{eq:ode:ci:generic:1}-\eqref{eq:ode:ci:generic:2}) reveals that direct-switch dynamics always yield a single stable equilibrium (Supplementary Text ST3). This equilibrium shifts toward either option depending on the interplay between asocial dynamics $\eta$, the quality ratio $q = q_a/q_b$, and asocial bias $\eta_a$. Even for relatively low $\eta$, the asocial bias $\eta_a$ can offset quality differences and lead to decision deadlocks or weak majorities. The critical bias $\eta_a^*$ (black line in the last row of Figure~\ref{fig:hmapci}) marks the point at which the majority opinion shifts between options (computed in the Supplementary Text ST4). Around this threshold, the system exhibits poor cohesion.

Cross-inhibition, by contrast, admits multiple equilibria that we study through the bifurcation diagrams of Figures~\ref{fig:density_synergetic} and~\ref{fig:density_antagonistic}. Direct-switch reaches a stable majority for the optimal option only when option qualities differ ($q \ne 1$) and asocial dynamics are rare (Figure~\ref{fig:density_synergetic}D for $\eta \le 0.05$ and Figure~\ref{fig:density_synergetic}G for $\eta \le 0.1$); otherwise, the swarm remains in an indecisive state, with a similar numbers of robots committed to either option. On the other hand, cross-inhibition maintains high cohesion across a wide range of parameters, consistent with previous findings~\cite{zakir2022,reina_cross-inhibition_2023}. Even under strong asocial dynamics ($\eta \lesssim 0.4$ for noise type-1 and $\eta \lesssim 0.17$ for noise type-2), cross-inhibition breaks symmetry and drives the swarm toward a clear majority, regardless of $q$ or $\eta_a$. This contrast with the indecision often seen with direct-switch is clearly visible in the cohesion heatmaps of Figure~\ref{fig:hmapds}.

\subsubsection{Accuracy}

Figure~\ref{fig:hmapds} also compares the two mechanisms in terms of accuracy, defined as the probability of reaching a large majority where at least $Q=0.75$ of the swarm selects the best option within $T=2\times 10^5$ time steps.
While both mechanisms can fail under extreme asocial biases, their failure modes differ.

Cross-inhibition is cohesive but may settle on the inferior option due to its bistable nature. This occurs when the swarm enters the basin of attraction for the suboptimal equilibrium, driven by initial conditions or stochastic fluctuations (see Supplementary Figure SF5). In contrast, direct-switch generally avoids selecting the inferior option when asocial dynamics are weak. However, under those same conditions, cohesion is low, and decisions are delayed. In fact, the apparent accuracy does not result from stable consensus but from transient fluctuations: the system hovers near an indecisive equilibrium, where small shifts in the majority can occasionally cross the quorum threshold $Q$ (see the robot distribution in Figures~\ref{fig:density_synergetic} and \ref{fig:density_antagonistic}). Because this equilibrium is biased toward the better option, such momentary majorities occur more frequently in its favour, yet the system remains attracted to the indecisive state, resulting in highly variable decision times (see Supplementary Figure SF4).

Notably, moderate levels of asocial dynamics can improve the accuracy of cross-inhibition by eliminating the suboptimal attractor, as previously observed~\cite{zakir2022,zakir_miscommunication_nodate}.
This effect is evident in the robot results (red histograms) in Figure~\ref{fig:density_synergetic}, where the lower suboptimal branch vanishes as $\eta$ increases, thereby improving decision accuracy (see also Supplementary Figure SF9D,F).
Importantly, this improvement is not merely a consequence of added noise (higher system temperature to escape from local minima); rather, asocial dynamics fundamentally reshape the system’s stability landscape. The bifurcation point $\eta^*$ marks the transition from bistability to monostability. At this critical point, the swarm achieves maximal accuracy by reliably selecting the better option; however, further increases of $\eta$ reduce group cohesion. Therefore, moderate asocial dynamics can improve decision accuracy by eliminating the possibility of selecting the inferior option.


\subsubsection{Speed}

Cross-inhibition consistently outperforms direct-switch in decision speed, both in reaching any decision and in selecting the better one (Figures~\ref{fig:hmapds}A and Supplementary Figure SF4). This speed advantage is particularly pronounced under antagonistic bias and remains robust to variations in the uncommitted duration $t_u$ (Supplementary Figure SF6). 
Combined speed--accuracy plots from robot simulations (Supplementary Figure SF11A-B) show a trade-off: cross-inhibition yields rapid, cohesive decisions, while direct-switch can achieve higher accuracy in some cases but often at the cost of weak majorities and delayed decisions.

\subsubsection{Scalability}

\begin{figure}[!ht]
  \centering
\includegraphics[width=0.69\textwidth,trim={0cm 0.1cm 0cm 0.5cm},clip]{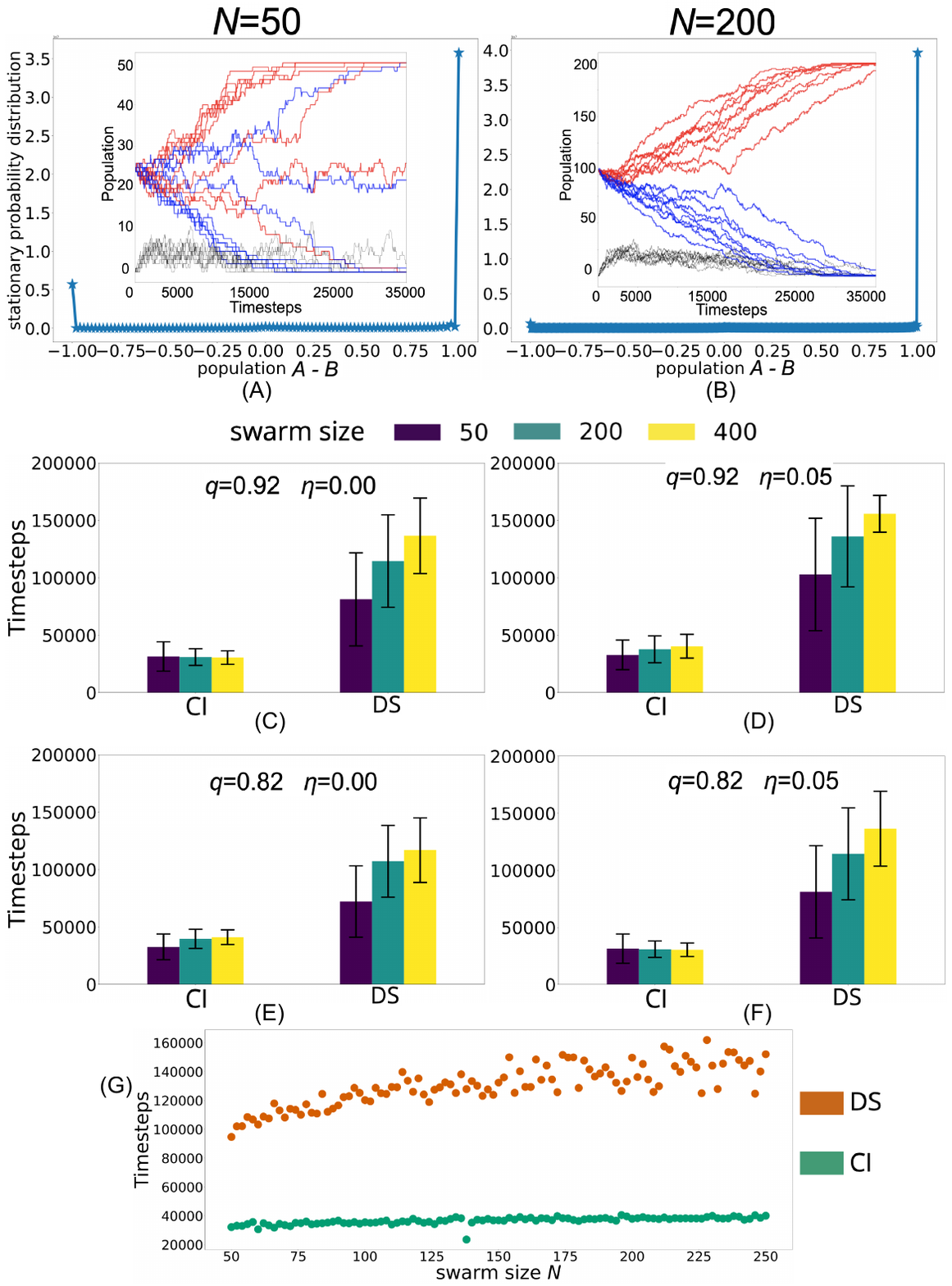}
      \caption{Effect of the swarm size $N$ on decision accuracy (A-B) and decision speed (C-F). Results from 250 Gillespie simulations. (A-B) Stationary probability distribution for cross-inhibition (type-1) for $N=50$ (A) and $N=200$ (B), when $q=0.82$, $\eta=0$, and the initial distribution of agents between options A and B is equal, i.e., $\{U, A_D, B_D, A_E, B_E\}=[0, 0.14N, 0.14N, 0.36N, 0.36N]$. The insets show the trajectories of 8 of the Gillespie simulation runs used to generate the SPD. The agents utilising cross-inhibition choose the inferior quality option less frequently in large swarms due to reduced stochastic fluctuations. (C-F) Time steps to make a collective decision (i.e., reach the quorum $Q=0.7$) for either option for both mechanisms, ci (cross-inhibition type-1) and ds (direct-switch) across three different $N\in\{50,200,400\}$ for $q\in\{0.92, 0.82\}$ and $\eta\in\{0,0.05\}$. (G) Scalability analysis for swarm size $N\in[50,250]$ for both cross-inhibition (type-1) and direct-switch. Cross-inhibition is faster than direct-switch to converge regardless of swarm size. This scalability analysis excludes physical interference, which can hinder decision-making at high robot densities \cite{hamann2022scalability}. We omit this aspect due to its strong dependence on specific robots, environments, and context-specific mitigation strategies.}  
  \label{fig:swarm_size}
 
\end{figure}

Beyond asocial dynamics, swarm size $N$ also influences collective decision outcomes~\cite{king2007bioletters}. 
Consistent with prior findings on direct-switch~\cite{valentini|ferrante|hamann|dorigo:2015}, we observe that larger swarms using cross-inhibition are more accurate (Figures~\ref{fig:swarm_size}A–B). This improvement is due to reduced stochastic fluctuations at larger $N$\cite{vankampen:1992}, which make it less likely for the swarm to settle on the suboptimal attractor when starting from two equally-sized subpopulations $A$ and $B$ (see trajectories in Figure~\ref{fig:swarm_size}'s insets).

Interestingly, decision speed under cross-inhibition remains nearly constant across swarm sizes (Figures~\ref{fig:swarm_size}C–G), while direct-switch becomes slower as $N$ increases.


\subsubsection{Comparison summary}

In summary, both direct-switch and cross-inhibition dynamics can be influenced by asocial dynamics $\eta$ and asocial bias $\eta_a$. When the goal is to reach consensus on the best option quickly, cross-inhibition generally performs better. It tends to be faster, more cohesive, and more scalable than direct-switch. Direct-switch may appear more accurate than cross-inhibition when selecting between options of similar quality. However, once asocial dynamics are present, direct-switch becomes less cohesive; consequently, its apparent accuracy often reflects weak majorities and delayed, fluctuation-driven quorum crossings rather than stable collective agreement~\cite{reina_cross-inhibition_2023}.

\subsection{Collective decision-making with more than two options}\label{sec:generalisability}

\begin{figure}[!ht]
  \centering
\includegraphics[width=0.93\textwidth,trim={0.5cm 8.5cm 0.5cm 1cm},clip]{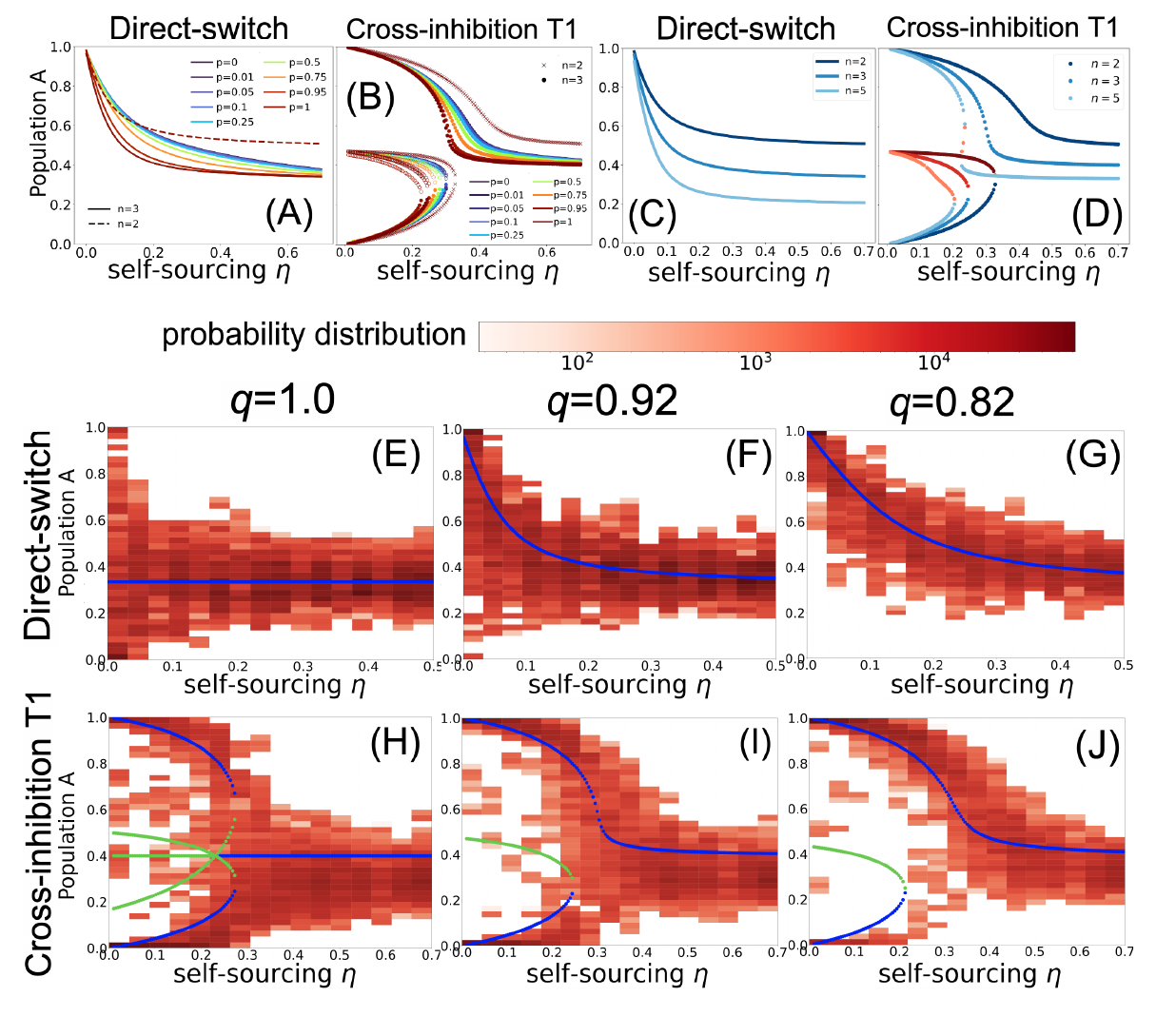}
  \caption{(A-D)~Equilibria of the basic ODE models showing the fraction of the population committed to option $A$ (y-axis) for direct-switch (A,C) and cross-inhibition type-1 (B,D) as a function of $\eta$ (x-axis), with $q=q_b/q_a=0.92$ and unbiased asocial dynamics distributed uniformly across the $n$ options ($1/n$ each). (A-B)~Binary ($n=2$) versus three-option ($n=3$) case for different values of $p\in[0,1]$, where $q_c=p\,q_b$. (C-D)~Comparison across $n\in\{2,3,5\}$ options for $p=1$, i.e., all inferior options have identical quality $q_i=0.92$ (with $q_a=1$). (E-J)~Robot simulation results and ODE bifurcation plots computed as in Figure~\ref{fig:density_synergetic} using the same parameters unless stated otherwise. We show the fraction committed to option $A$ (y-axis) in the collective-perception scenario with $n=3$, unbiased asocial dynamics $\eta_a=\eta_b=\eta_c=1/3$, $p=1$ (i.e., $q_b=q_c$) and quality ratios $q\in\{1,0.92,0.82\}$. Panel~H shows three adjacent unstable branches because we plot a one-dimensional projection (population A on the y-axis) of a three-dimensional dynamical system. A 3D view showing the distinct unstable eigendirections is provided in the supplementary videos. Robot simulations closely match the ODE predictions, and the qualitative trends from the binary case generalise to $n>2$: cross-inhibition continues to break symmetry and maintain cohesive decisions across broad conditions, whereas direct-switch increasingly exhibits weak majorities or deadlock as $\eta$ or $n$ increases.}
  \label{fig:fign}
\end{figure}

While the previous analysis in the binary setting facilitated interpretation of dynamics and results, we next test whether these findings generalise to collective decisions among more than two options (i.e., $n>2$).
We first consider the case $n=3$ and study how the dynamics change as $q_c$, the quality of the third option C, varies relative to $q_b$. Specifically, we set $q_c=p\,q_b$, where $p\in[0,1]$, so that $q_a \ge q_b \ge q_c$. Figures~\ref{fig:fign}A-B show how the stable equilibria of the basic ODE models (Equations SE13 and SE14) vary as a function of $\eta$ for different values of $p$, under unbiased asocial dynamics ($\eta_a=\eta_b=\eta_c=1/3$). Compared with the binary ($n=2$) case, collective decisions among three options are generally more difficult and therefore lead to lower cohesion. One exception occurs for low values of $p\ll1$ and moderate asocial dynamics, where direct-switch reaches slightly higher cohesion than in the binary case. In this regime, a small fraction of agents (approximately $\eta/3$) commit to option C through asocial dynamics and, because $q_c$ is low, communicate their opinion only infrequently. This behaviour is reminiscent of the uncommitted state in cross-inhibition, and may explain why direct-switch achieves slightly higher cohesion in the $n=3$ case than in the binary setting. This effect vanishes for higher values of $p$ or $\eta$. In general, $p=1$ is the most challenging decision setting, while lower values of $p$ make the dynamics progressively more similar to the binary setting. We therefore focus the next analysis on $p=1$.

Figures~\ref{fig:fign}C-D show that, as the number of options increases ($n\in\{2,3,5\}$), direct-switch undergoes a marked reduction in cohesion, even for relatively small levels of asocial dynamics ($\eta < 0.1$). By contrast, cross-inhibition maintains comparable cohesion for low asocial dynamics ($\eta < 0.15$), but transitions to decision deadlock at progressively lower values of $\eta$ as $n$ increases. This reflects the particularly challenging unbiased case, where asocial bias is distributed equally across all $n$ options, so that the probability of an asocial update favouring the best option decreases as $n$ increases ($\eta_a=1/n$). Analyses under other bias configurations (Supplementary Figure SF12, for different values of $\eta_a$ and $q$) show that direct-switch rarely reaches a sizeable majority for $n>3$ once asocial dynamics are present, even at low levels. Cross-inhibition, in contrast, always breaks symmetry and achieves high cohesion, although sometimes in favour of an inferior option, as already observed in the binary setting. These ODE results are in good agreement with the robot simulations, shown as 2D histograms in Figures~\ref{fig:fign}E-J.

\begin{figure}[p]
  \centering
\includegraphics[height=0.75\textwidth,trim={0cm 0cm 0cm 0cm},clip]{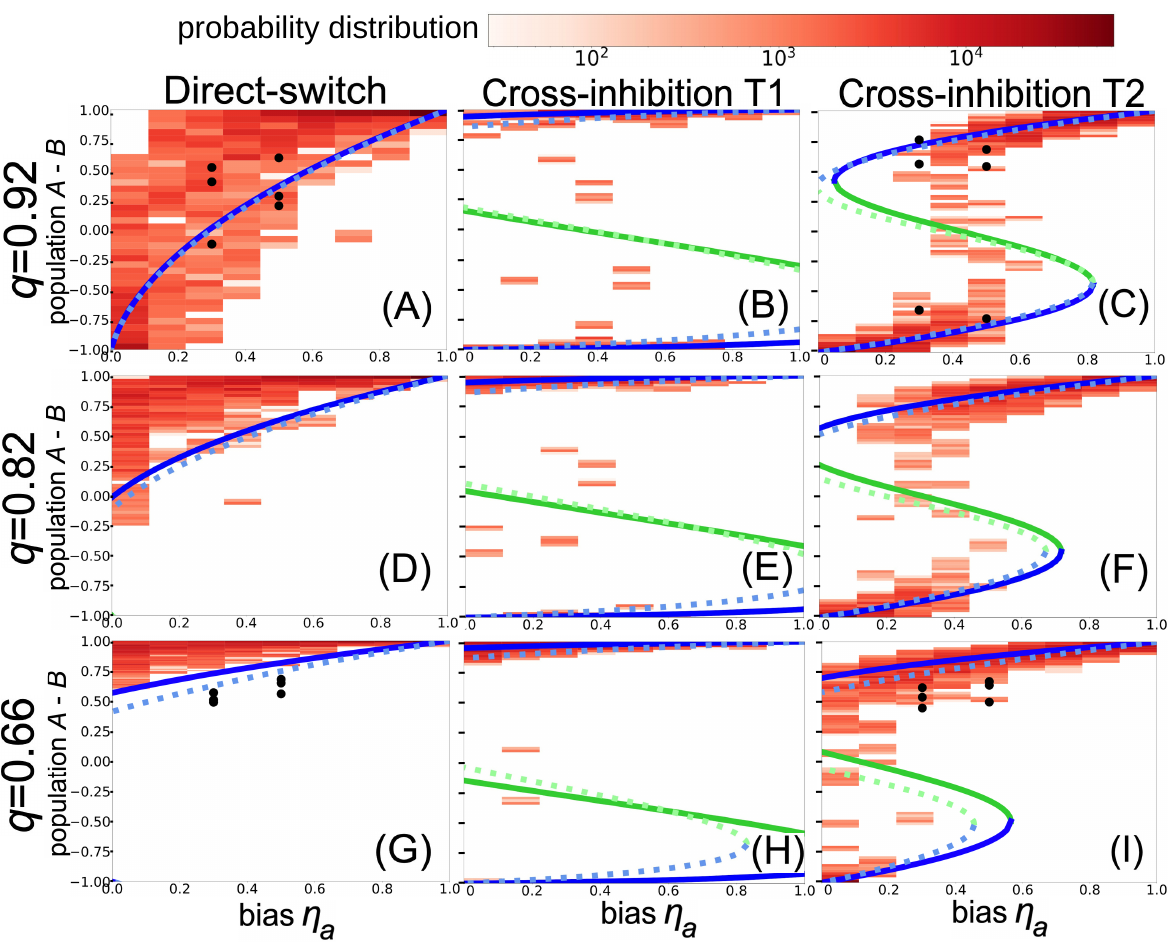}
\caption{Results from robot simulation, real-robot experiments, and ODE bifurcation analysis, computed as in Figure~\ref{fig:density_synergetic} using the same parameters unless stated otherwise. Here, we show the difference between the fractions of robots committed to options A and B ($A-B$, y-axis) as a function of self-sourcing bias towards option A ($\eta_a$, x-axis), with $\eta=0.05$. Values of $\eta_a<0.5$ correspond to antagonistic bias, $\eta_a>0.5$ to synergistic bias, and $\eta_a=0.5$ to unbiased asocial dynamics. Black dots in panels A, C, G, and I show the final fractions of robots from selected real-robot experiments (see Methods). When the two options are moderately similar (A,D), direct-switch establishes a majority only over a limited range of bias values $\eta_a$. In panel A ($q=0.92$), for bias values of approximately $\eta_a\approx 0.25$, the two subpopulations are equally sized ($A-B \approx 0$). For large quality differences (low $q$) or strong bias toward the best option (high $\eta_a$), direct-switch leads to a majority for the best option, although in certain condition can be relatively slim (e.g., high antagonistic bias $\eta_a \ll 0.5$ in D). By contrast, cross-inhibition never goes into a state of deadlock and always breaks the symmetry across the full range of $\eta_a$, but may occasionally select the inferior quality option.}
  \label{fig:density_antagonistic}
\end{figure}

\begin{figure}[!ht]
  \centering
\includegraphics[width=0.59\textwidth,trim={0cm 0cm 0cm 0cm},clip]{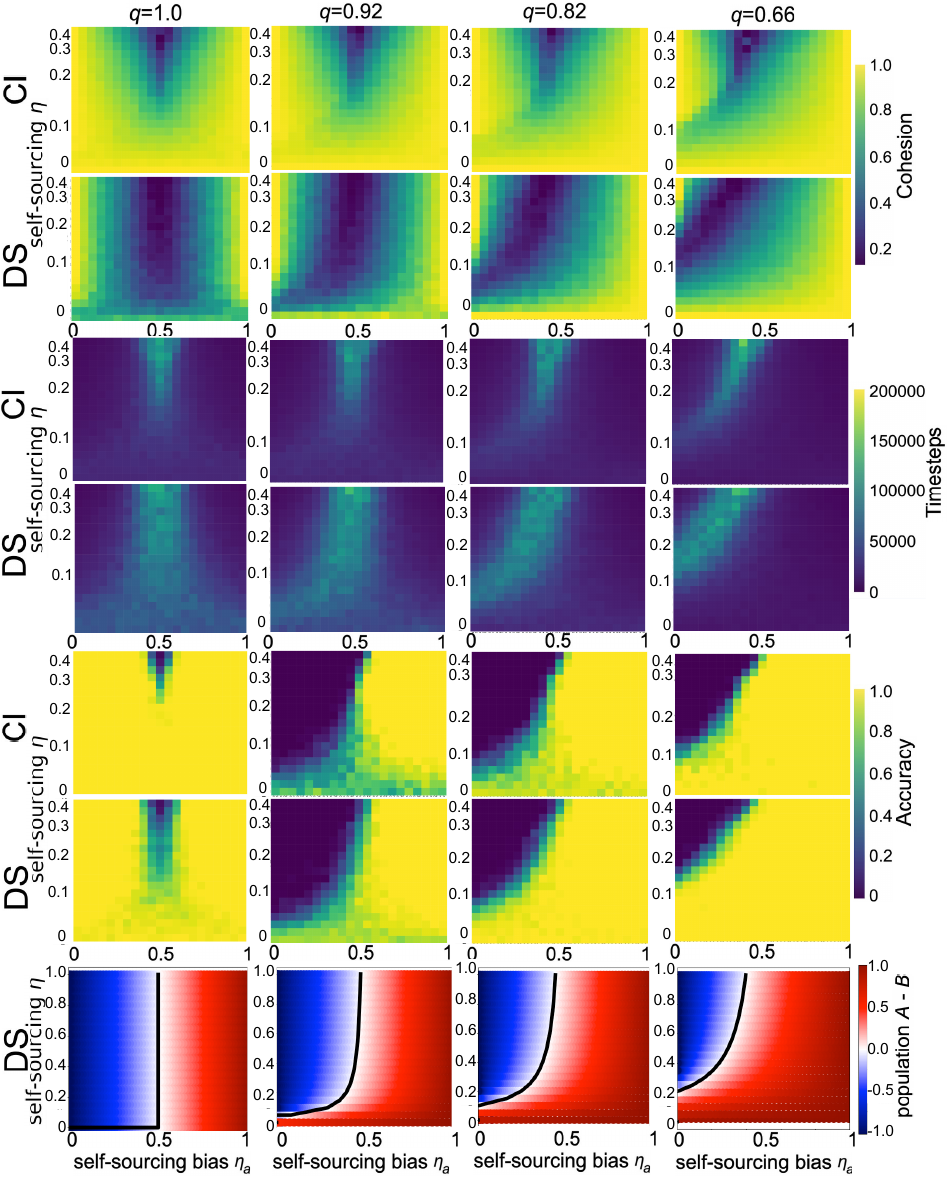}
  \caption{Comparison of the direct-switch (DS) and cross-inhibition (CI) type-2 mechanisms in terms of decision cohesion, speed, and accuracy as functions of asocial dynamics $\eta$ (y-axis) and asocial bias towards option A, $\eta_{a}$ (x-axis). Values $\eta_a>0.5$ correspond to synergistic bias, whereas $\eta_a<0.5$ to antagonistic bias favouring the lower-quality option B. Heatmaps were generated from 100 Gillespie simulations of $T=200\,000$ time steps. Parameters $t_d$, $t_e$, and $t_u$ are the same as those used in the ODE analysis and robot experiments presented in Section~\textit{\nameref{sec:3robot}}. Results are shown for a swarm of $N=100$ agents, initialised in state $[A_D, B_D, A_E, B_E]=[14,14,36,36]$ and for quality ratios $q=\{1,0.92, 0.82,0.66\}$. The first two rows compare group cohesion and show that swarms using cross-inhibition are consistently more cohesive. The second two rows compare decision speed and show that cross-inhibition is faster than direct-switch across most of the parameter space. The speed's standard deviation is reported in Supplementary Figure SF7. The third two rows compare decision accuracy. Direct-switch shows higher accuracy than cross-inhibition in the bottom-left region when the decision problem is easy ($q > 0.92$); however, this region is also associated with relatively low cohesion. The last row shows the long-term population difference (number of robots in state $A$ - number of robots in state $B$) / $N$, obtained from the equilibrium of the direct-switch ODE system SE1. Red indicates convergence toward option A; blue toward option B. The black line marks the maximum-deadlock point ${\eta_a}^{*}$ where $A=B$, computed in Supplementary Eq. SE5 as a non-linear function of $\eta$ (y-axis), $\eta_a$ (x-axis), and $q$. }
  \label{fig:hmapds} \label{fig:hmapci}
  
\end{figure}

\section{Discussion}

Minimal mechanisms of social agreement allow groups of simple agents to compensate for individual errors and collectively make fast, accurate, and cohesive decisions. 
Our analysis reveals that the decision performance can be dramatically altered by the presence of asocial dynamics, which systematically bias the flow of social information.
Despite, or thanks to, its simplicity, the biologically grounded mechanism---cross-inhibition---exhibits remarkable robustness.

Previous studies have analysed the dynamics of such mechanisms either under idealised, bias-free conditions \cite{valentini|hamann|dorigo:2014,valentini|ferrante|hamann|dorigo:2015} or with biases specific to the application scenario~\cite{talamaliSciRob2021,Prasetyo2019,zakir2022,leaf2024}. Our mathematical models unify a broad class of asocial dynamics---covering zealots, corrupted communication, and self-sourcing---by representing them as equivalent terms in the opinion-update equations, enabling systematic and exhaustive analysis of their effects on collective decision-making.
Self-sourcing can be necessary for adaptation to environmental changes~\cite{talamaliSciRob2021,salahshourphyrev2019}, while the presence of zealots or corrupted communication may be unavoidable due to faults or cyberattacks \cite{Canciani2019,AntZakDorRei2024:aamas,giulia2021}.
These challenges are not unique to artificial systems. 
Natural systems must also resolve internal conflicts, as in animal groups with informed individuals pursuing opposing goals~\cite{couzin2011,Rajendran2022}, or human collectives where misinformation can shape decision outcomes~\cite{coleman2021,Centola2018}. This mean-field equivalence is inherently macroscopic: it captures how asocial influences bias population-level dynamics, but may mask implementation-specific asymmetries that become significant in finite or spatially structured swarms. Accordingly, analogies between robotic and natural systems should be interpreted primarily in qualitative terms. Nevertheless, the unified formulation helps identify key ingredients---such as inhibitory interactions---that can reshape collective dynamics and improve robustness under biased conditions.

While direct-switch is typically more accurate than cross-inhibition in the absence of asocial dynamics (Figure~\ref{fig:density_synergetic} and SF9), our analysis shows that even mild asocial influence often drives the swarm into a state of decision deadlock. Cross-inhibition, a biologically inspired mechanism qualitatively equivalent to decision-making models in house-hunting honeybees~\cite{seeley|etal:2012} and neural circuits~\cite{PIRRONE202266,usher2001,Borofsky2020}, enables faster, more cohesive, and more scalable decisions under a wide range of biased conditions. Despite strong asocial dynamics, robot swarms using cross-inhibition consistently outperformed those using direct-switch in both binary and multi-option settings. While natural and artificial systems often approximate multi-alternative decisions as sequences of binary comparisons~\cite{Sridhar2021,OddiANTS2022,LeeANTS2018}, we also analyse best-of-$n$ dynamics directly for $n>2$ and recover the same qualitative trends as in the binary case. More broadly, decision-making is known to slow as the number of alternatives increases (Hick–Hyman's law~\cite{Reina:scirep:2018}), making mechanisms that reliably break symmetry and maintain cohesion increasingly relevant in multi-option settings. Crucially, experiments with 50 physical robots reproduce the qualitative trends predicted by the models and simulations, supporting the persistence of these effects under embodied non-idealities such as errors in motion and communication. Our results reveal a previously unrecognised robustness of cross-inhibition, which may explain its recurrence across biological systems.

Cross-inhibition has been linked to better management of the speed-value trade-off~\cite{pais|etal:2013,Reina:DARS:2016}, where decision value reflects the reward obtained from the selected option. By enabling consensus even among similar options or under strong bias, cross-inhibition can occasionally lead to the selection of a suboptimal alternative. Natural systems such as brains and insect colonies may favour coherence and speed over accuracy, preferring a rapid, unified choice over prolonged indecision~\cite{PIRRONE202266}.
For example, honeybees may have evolved the stop signal to prevent swarm splitting during nest-site selection~\cite{seeley|etal:2012,Reina:PRE:2017}. In contrast, ants lack such inhibitory signalling~\cite{franks|pratt|mallon|britton|sumpter:2002}, possibly because split colonies can reunite days after relocation~\cite{franks|etal:2013}. Inhibitory signalling between competing integrators of noisy evidence also enables winner-take-all dynamics in other domains, from mitotic checkpoint control~\cite{Cardelli2012,Cardelli2017} to neural decision-making via leaky competing accumulator models~\cite{PIRRONE202266,usher2001}. The recurrence of cross-inhibition across such diverse systems suggests it may be a particularly robust and scalable solution for fast, cohesive decision-making.

Our analysis reveals that cross-inhibition is not only robust to asocial biases but can also be antifragile to them. Antifragility~\cite{Taleb2013} refers to the property of systems that improve when exposed to moderate stressors. In collective decision-making, this is exemplified by systems that perform better under moderate levels of miscommunication~\cite{salahshourphyrev2019,zakir_miscommunication_nodate} or balanced groups of stubborn agents (zealots)~\cite{JuliaKlein2024,Prasetyo2019}. Our bifurcation analysis shows that the performance gains from asocial dynamics are not solely attributable to increased randomness (system temperature) but also to structural changes in the system’s phase space. Performance peaks near the edge of criticality, where the system balances flexibility and stability---a regime that has also been shown to enhance responsiveness and adaptability in other collective systems that require agile behaviour\cite{leonard2024fast,gomez2023fish}. Since asocial dynamics may be inherent in natural or engineered systems, mechanisms that benefit from such disturbances---rather than merely resist them---offer a powerful design paradigm. Cross-inhibition, inspired by biological systems, exemplifies this potential.

Our analysis clarifies the trade-offs of robot swarms running algorithms based on either mechanism. Under asocial dynamics, direct-switch and cross-inhibition reveal distinct vulnerabilities analogous to denial-of-service (DoS) and wrong-addressing attacks---concepts borrowed from cybersecurity. DoS, where a system is overwhelmed and rendered unresponsive, mirrors how direct-switch can become locked in indecision under antagonistic bias. Conversely, wrong-addressing, akin to astroturfing that skews perception toward a misleading outcome, parallels cross-inhibition’s tendency to select a suboptimal option when qualities are similar. These analogies highlight distinct failure modes and help characterise the specific risks associated with each decision-making mechanism.

Certain application scenarios may benefit from the dynamics of direct-switch. When robot opinions are reported to human operators, prolonged indecision or slim majorities can serve as valuable indicators of ambiguous or difficult choices between similarly valued options. In certain cases, delaying action may even be advantageous, allowing the system to remain inactive until clearer information emerges or until acting becomes safe.

By contrast, cross-inhibition is better suited for autonomous swarm systems that must act decisively without human intervention. It enables rapid consensus even when option qualities are close, favouring operational responsiveness over guaranteed optimality. This makes it well-suited for tasks where timely action is critical, for example, swarms tasked with identifying the agricultural field most in need of treatment~\cite{dario2017}, selecting the first chemical spill to contain, or prioritising the most urgent firefighting front~\cite{Tzoumas2024SI}.

We investigated two key decision-making mechanisms in their minimal form. While combining social and asocial information is optimal in several biological collectives~\cite{Nagy2020,Sosna2019}, the relative weighting of these sources critically shapes the group dynamics~\cite{Reina:PRE:2017,salahshourphyrev2019}.
Our findings motivate algorithms in which robots dynamically adjust the balance between social and asocial inputs to steer the system toward criticality, a condition that, as our analysis suggests, enhances collective decision-making performance. A key challenge lies in the limited information available to individual agents, which must rely on local cues to regulate their behaviour, for example, estimated local consensus~\cite{march2024honeybee}, interaction frequency \cite{Rausch:SwInt:2019}, or urgency of action~\cite{talamali2019improving}.
Future strategies could hybridise direct-switch and cross-inhibition---e.g., by incorporating periodic opinion resets to escape local minima~\cite{leonard2024fast}---or address more complex asocial behaviours, such as contrarians with population-dependent biases~\cite{Canciani2019,zakir_miscommunication_nodate}. Finally, because the structure of the interaction network influences group decision-making~\cite{Becker2017,Sosna2019,Reina:PRE:2024}: dynamically adjusting network topology may provide individuals with additional leverage to steer collective dynamics or mitigate asocial biases~\cite{talamaliSciRob2021}.

\section*{Methods} 

\subsection{Mean-field models of the basic collective decision-making algorithm}\label{sec:basic-ODE}

The opinion dynamics of the swarm can be modelled using mean-field models in the form of a system of ordinary differential equations (ODEs). This model assumes the continuous limit $N \rightarrow \infty$, that is, the swarm has infinite size (finite-sized system models are discussed in the  Section~\textit{\nameref{sec:stochastic_analysis}}). This assumption is effective for making good approximations of large-scale, well-mixed systems and enables the use of dynamical systems and bifurcation analysis to generate deterministic predictions of the system's behaviour \cite{strogatz2018,leonard2024fast}. 

We focus on the best-of-$n$ decision problem, where a swarm of $N$ agents---or robots---exchange information to select the best option collectively. We present the binary models ($n=2$ options) here, and discuss the general $n$-option formulation in Supplementary Text ST6. Options A and B are associated with a quality value, $q_a$ and $q_b$, respectively. The ratio of the two qualities is $q=q_b/q_a$. Without loss of generality, we assume that $q_a \ge q_b >0$, which implies that $q\in(0,1]$. How an agent committed to A (in state $A$), committed to B (in state $B$), or uncommitted (state $U$) changes its opinion depends on the opinion update mechanism. We consider two mechanisms: direct-switch and cross-inhibition (see Figure~\ref{fig:robot-fsm}A-B).

Let $a, b$, and $u$ be the proportions of robots in the states $A, B$, and $U$, respectively. Note that, in these basic models, we do not consider separated dissemination and exploration phases (which are only relevant to the robot-specific models, presented in the Supplementary Equations SE1-SE4).

The ODE system for the direct-switch model with generic asocial dynamics reads as
\begin{align}
  \label{eq:ode:vm:generic}
    \od{a}{t}= a b (1-\eta) (q_a  - q_b) + \eta (b \eta_a-a \eta_b)
\end{align}
where $b=1-a$.

The ODE system for the cross-inhibition type-1 model with generic asocial dynamics reads as
\begin{align}
  \label{eq:ode:ci:generic:1}
  \od{a}{t}= a (1-\eta)  (q_a u -q_b b ) + \eta (u \eta_a-a \eta_b)
  \\
    \label{eq:ode:ci:generic:2}
  \od{b}{t}= b (1-\eta)  ( q_b u - q_a a ) + \eta (u \eta_b-b \eta_a)
\end{align}
where $u=1-a-b$. The cross-inhibition type-2 model is presented in Supplementary Equation SE3.

In the direct-switch model, the proportions $a$ and $b$ increase or decrease when a robot in state $A$ (or $B$) directly switches to $B$ (or $A$) after interacting with a robot committed to the opposite opinion, namely following a social interaction occurring with probability $1-\eta$. This interaction occurs with a frequency proportional to the quality of the option held by the neighbour, i.e., $q_a$ or $q_b$. The proportions of robots $a$ and $b$ also change due to asocial dynamics occurring at rate $\eta$. A robot in state $A$ (or $B$) switches to $B$ (or $A$) at a rate $\eta \eta_b$ (or $\eta \eta_a$).

Similarly, in cross-inhibition, an uncommitted robot $U$ is recruited to option A or B after interacting with a committed robot, also occurring at rates proportional to $q_a$ or $q_b$, respectively; once again, this social interaction occurs with probability $1-\eta$. Proportions $a$ and $b$ also decrease at rates $(1-\eta)q_b$ and $(1-\eta)q_a$, respectively, when interaction among robots committed to different options occurs. An uncommitted robot---in state $U$---switches to $A$ (or $B$) at a rate $\eta \eta_a$ (or $\eta \eta_b$), and committed robots in state $A$ (or $B$) become uncommitted at rate $\eta \eta_b$ (or $\eta \eta_a$). 

In both models, when $\eta_{a} > \eta_{b}$, asocial dynamics are synergistic, while when $\eta_{a} < \eta_{b}$, they are antagonistic.

\subsection{Equivalence of asocial dynamics}\label{sec:equivalence}

A robot updates its opinion by polling social information from neighbours with probability $\eta$, but can also be influenced by asocial dynamics such as messages from zealots, corrupted signals, or independently sourced opinions with probability $1-\eta$. The asocial dynamics can be biased towards either option, where $\eta_a$ is the probability that asocial updates provide information in favour of A, and $\eta_b$ in favour of B. More concretely, when $z$ zealots are present, $\eta=z/N$ is the probability of interacting with a zealot, and $\eta_a = z_a/z$, $\eta_b = z_b/z$ represent the proportions of zealots favouring each option ($z_a + z_b = z$). In the case of MITM attacks, $\eta$ is the probability of receiving a corrupted message, while $\eta_a$ and $\eta_b$ capture how often messages are altered to support option A or B. For self-sourcing, $\eta$ is the probability that a robot uses environmental information rather than social input, and $\eta_a$, $\eta_b$ represent how likely the robot is to independently discover evidence in favour of A or B---that is, how easily each option can be found in the environment. Although the interpretations differ across these cases, all three events involve updating the robot's opinion based on biased information rather than representative social input from the population.

This unified formulation simplifies the analysis of the mean-field models and enables systematic exploration of the parameter space defined by $\eta$ and $\eta_a$. In the ODE models, opinion changes are the result from two concurrent processes: the decision-making mechanisms---defined by direct-switch (weighted voter model) or cross-inhibition---and biased asocial events, which act as a constant biased `opinion field'. This abstraction allows experimental results to generalise across contexts: the same effect can be achieved by modifying the number of zealots, altering the rate of message corruption, or adjusting environmental features that affect independent discovery of options (see Supplementary Material ST5).

\subsection{Collective perception through robot-specific collective decision-making algorithms}\label{sec:dec_strategy}

The basic algorithms shown in Figures~\ref{fig:robot-fsm}A-B can be instantiated to make best-of-$n$ decisions in a robot swarm. Robots are physical devices that operate within the technical constraints of the platform and must sense and sample the environment to determine the quality of options. Therefore, we extend the basic decision-making algorithms by including two alternating phases: exploration and dissemination. 

The robot behaviour for the binary ($n=2$) case is described by the finite state machine shown in Figure~\ref{fig:robot-fsm}C. Robots can be in five possible states: $A_D$ (disseminating opinion A), $B_D$ (disseminating opinion B), $A_E$ (exploring option A), $B_E$ (exploring option B) and $U$ (uncommitted to any option). Robots in the exploration phase sample the environment and estimate the quality of their current opinion, i.e., option A or B. The time spent in the exploration state is drawn from an exponential distribution with rate $\lambda_e = t_e^{-1}$. Therefore, on average, robots remain in the exploration for a time $t_e$ before transitioning to the dissemination state $A_D$ or $B_D$. Robots in the dissemination state locally share their opinions (i.e., option A or B) with neighbouring robots. The time spent in the dissemination state is drawn from an exponential distribution with rate $\lambda_d=(q_i t_d)^{-1}$. Consequently, the mean time spent in the dissemination state is proportional to the assessed quality $q_i$ (where $i= \{a, b\}$) of the option evaluated during the exploration phase. Uncommitted robots stay in the uncommitted state $U$ for a mean time $t_u$, during which they neither assess nor disseminate any option.

At the end of the dissemination state or the uncommitted state, the robot updates its opinion by polling social information. In the default case, this information is social, that is, sampled from neighbouring robots. This polling phase can also be influenced by asocial dynamics caused by zealots, MITM attacks, or independent self-sourcing of information from the environment (Figure~\ref{fig:bias}F and Supplementary Figure SF8). Although these mechanisms differ in origin, they lead to the same probabilistic update process. With probability $1 - \eta$, the robot polls representative social information from one of its neighbours, meaning that the probability of observing opinion A or B reflects the population-level bias towards one option or the other. With probability $\eta$, the robot uses asocial information, which may be broadcast by a zealot robot, corrupted via an MITM attack, or independently sourced from the environment. In the latter case, strictly speaking, the robot does not poll information but instead spontaneously decides not to use social information and updates its opinion by sampling the environment. In this work, we focus our analysis on independently sourced environmental information.

Depending on the outcome of the polling action (or self-sourcing), the robot returns to an exploration state $A_E$ or $B_E$, or to the uncommitted state $U$ (Figure~\ref{fig:bias}F and Supplementary Figure SF8). 

There are two main differences between this robot algorithm and the basic one: first, the decision-making dynamics are slower because of the additional dissemination state; second, only a fraction of committed robots broadcast messages. When, with probability $1 - \eta$, a robot polls representative social information (right branch in Supplementary Figure SF8), it polls information only from robots in disseminating states, that is, $A_D$ and $B_D$. Hence, the probability of receiving a message supporting option A is $A_D/(A_D+B_D)$, while the probability of receiving a message supporting option B is $B_D/(A_D+B_D)=1- [A_D/(A_D+B_D)]$. These differences lead to slight changes in the dynamics and equilibria, while yielding qualitatively similar results (Figures~\ref{fig:density_synergetic} and~\ref{fig:density_antagonistic}). The ODEs modelling the robot simulations are given in Supplementary Text ST2.

\subsection{Stochastic Analysis}\label{sec:stochastic_analysis}
In Supplementary Text ST2, we modelled the behaviour of the two collective decision-making algorithms by using the continuous-limit approximation, where the swarm size $N$ approaches infinity. However, real-world swarm systems comprise a large yet finite number of robots. This finite system size can cause random, size-dependent fluctuations that may significantly affect the system dynamics~\cite{toral|tessone:2007,biancalani2014}. Therefore, it is always important to verify whether the predictions derived from continuous approximations remain valid in the presence of size-dependent fluctuations. 
To study finite-size effects in a computationally efficient way, we use the formalism of chemical master equations derived from a chemical-like reaction network~\cite{vankampen:1992,gillespie2013perspective}. Chemical master equations describe the stochastic time evolution of the probability distribution over system states generated by coupled reactions. In a multi-agent system, agent states are represented by molecule species, and state transitions are represented as chemical reactions with specific rates. The transitions for the algorithms based on direct-switch and cross-inhibition mechanisms in our system are specified in Supplementary Text ST7.

The chemical reaction network and corresponding master equation are advantageous because they allow the study of random fluctuations with a principled approach rather than introducing noise with an arbitrarily chosen magnitude. In this formalism~\cite{vankampen:1992}, the magnitude of the fluctuations is inversely proportional to system size $N$. This means that, as the swarm size $N$ increases, these fluctuations become smaller, and in turn, the predictions of the chemical reaction network converge to those of the noiseless ODE model. Solving chemical master equations analytically can be challenging and, in some cases, intractable. We therefore investigate these systems numerically using the Gillespie algorithm \cite{gillespie2013perspective}. Unless stated otherwise, the analyses based on the Gillespie algorithm use 50 simulations of 20\,000 time steps each. These simulations use the same transition rates as the ODE models and robot simulations and experiments, with each run initialised using random initial conditions where the number of robots in each state is drawn uniformly at random.

\subsection{Robot experiments}\label{sec:3robot}

To perform robot simulations and experiments, we consider a collective perception case study~\cite{zakir2022,Hamann:ANTS:2026} in which the robots are tasked with selecting the most important element in the environment. In our case, the floor is composed of randomly distributed red and blue tiles (see Figure~\ref{fig:bias}B and SF10), and the swarm must collectively select one of the two colours. Therefore, the two colours are the available options: A is red, and B is blue.

\subsubsection*{Synergistic and antagonistic environmental biases}
The values $\eta_a$ and $\eta_b$ indicate the proportion of tiles allocated to each option (red and blue, respectively), and thus define the probability of discovering that option. When we consider synergistic asocial dynamics, the proportion of tiles of each colour is linearly proportional to the quality of the corresponding option, i.e., $\eta_a=q_a/(q_a+q_b)$ and $\eta_b=1-\eta_a=q_b/(q_a+q_b)$. More precisely, the abundance of each colour represents its quality~\cite{valentini|brambilla|hamann|dorigo:2016,ebert2018multi,palina2019}. Therefore, to estimate option quality and collectively select the most frequent colour in the environment, robots count the proportion of tiles of each colour.

Under antagonistic asocial dynamics, the higher-quality option is less likely to be self-sourced ($q_a>q_b$, while $\eta_a<\eta_b$). This scenario corresponds to the cases where robots discover lower-quality options more easily than the best option, which is more difficult to find~\cite{reina|valentini|fernandezoto|dorigo|trianni:2015}. The proportion of tiles of each colour only determines the bias in the discovery of either option (i.e., robots find the more abundant colour more often). Instead, quality is not represented by colour abundance but is an independent quantity that robots estimate from the environment.

\subsubsection*{Simulation setup} 
We simulate a swarm of $N=100$ robots modelled after the Kilobot platform~\cite{rubenstein|ahler|hoff|cabrera|nagpal:2014}. The Kilobots are low-cost, small-sized robots that move at a speed of \unitfrac[1]{cm}{s}, rotate in place at \unitfrac[45]{$^\circ$}{s}, communicate with each other over a range of \unit[10]{cm} via infrared (IR), and operate with a control loop interval of approximately \unit[32]{ms}. We use a simulation of the Kilogrid system~\cite{ValAntTra-etal2018:si}, a customisable and interactive environment specifically tailored for the Kilobots. The Kilogrid, which comprises 800 cells, allows robots to self-source options via IR communication and assess the quality of their opinions during exploration. 

In the case of synergistic bias, the proportion of cells of each colour represents both the option quality and the bias in the asocial dynamics. Therefore, when $q_a = q_b$, 50\% of the tiles are red and 50\% are blue, and when $q_a \neq q_b$, the ratio between $q_a$ and $q_b$ defines the proportion of cells of each colour. In our experiments, we test three tile proportions in the Kilogrid:  50\%A-50\%B ($q=1$, no quality difference), 52\%A-48\%B ($q=0.92$, small quality difference), and 55\%A-45\%B ($q=0.82$, moderate quality difference).

In the case of antagonistic bias, the quality is a number in the range $q_i\in[0,1]$ communicated by the Kilogrid.
Given a percentage of red and blue cells, their spatial distribution in the Kilogrid is uniformly random. All cells, except those at the borders (shown in Figure~\ref{fig:robot-fsm}E and Supplementary Figure SF10), are programmed to continuously transmit IR messages containing their ID, their colour, and, in the antagonistic case study, the quality value $q_i$. Both the Kilobots and the Kilogrid are simulated in ARGoS through dedicated plugins \cite{Pinciroli2018,TillANTS2022}.

\subsubsection*{Initialisation}
At the beginning of each run, we initialise the robots in the exploration state with a random initial opinion, with half of the swarm committed to option A (in state $A_E$), and the other half committed to option B (in state $B_E$). We deploy the robots on the Kilogrid at uniformly random initial positions. The distribution of Kilogrid tiles is randomly regenerated for each run. Each run lasts $T=200\,000$ simulation time steps, corresponding to 110 minutes.

\subsubsection*{Robot behaviour} 
To explore different areas of the environment and interact with different robots, the Kilobots perform a random walk. This enables them to encounter different neighbours during the dissemination phase and to estimate option quality more accurately during the exploration phase. 
Due to the absence of proximity sensors, Kilobots rely on the Kilogrid to detect nearby walls. Specifically, Kilogrid cells transmit a binary `wall flag' (high or low) to signal proximity to a wall. Cells at the borders and cells adjacent to them transmit a high wall flag, while all other internal cells transmit a low flag. When a robot receives a high wall flag from the Kilogrid, it executes a basic obstacle-avoidance routine regardless of its current state. 

The robot behaviour during the exploration phase is different in the synergistic and antagonistic cases. In the synergistic case, a robot committed to option $i$ reads Kilogrid messages to track the total number of cells encountered, denoted by $T_c$, and the number of visited cells whose colour matches its opinion, denoted by $C_i$. The robot ensures that each cell is counted only once by using the cell's ID. At the end of the exploration cycle, the robot estimates the option quality as $q_i= C_i/T_c$. The values of $T_c$ and $C_i$ are specific to a single exploration cycle and are reset before entering the dissemination state. If the robot does not visit any cell corresponding to option $i$, it sets the estimated quality to $q_i=0$. In the antagonistic case, the robot immediately estimates $q_i$ upon visiting a cell whose colour matches its opinion $i$. The quality estimate communicated by the cell is sampled from a normal distribution $\mathcal{N}(q_i, \sigma)$, where $q_i$ is the nominal quality value, either $q_a$ or $q_b$, and $\sigma$ is the sampling noise. The value of $\sigma$ is set by computing the standard deviation of the noisy quality estimates the robots obtain from the Kilogrid, relative to the nominal quality values (Figure~\ref{fig:robot-fsm}D). To keep the two cases as comparable as possible, the mean duration of the exploration phase is the same in both the synergistic and antagonistic cases. 
Based on $q_i$, the robot determines the dissemination time using an exponential distribution with mean $\lambda^{-1}_d = q_i t_d$, where $t_d$ denotes the average number of control cycles spent in dissemination when $q_i=1$, and is set to 1\,300 in our setup. This value corresponds to a mean dissemination time of approximately 40 seconds. 
Uncommitted robots move randomly through the environment for an average of $t_u=1\,000$ control cycles (approximately 30 seconds), during which they neither estimate quality nor transmit messages. Then, they transition to the polling or self-sourcing phase.

After dissemination, the robot determines with probability $\eta$ whether to update its opinion by self-sourcing information from the environment or by engaging in social interaction. If the robot chooses to self-source, it switches its opinion to the option (colour) of the cell at its current position. Therefore, the proportions of tiles of each colour, $\eta_a$ and $\eta_b$, determine the probabilities of switching opinion to A or B, respectively.
In the synergistic case, $\eta_a=q_a /(q_a+q_b)$, hence switching to state $A_E$, in favour of the higher-quality option A, is more likely than switching to $B_E$. In the antagonistic case, we assume $q_a>q_b$, hence $\eta_a<\eta_b$.

Given the robots' extremely limited memory and computation, during the polling phase each robot only considers the first message it receives from its neighbours and uses it to update its opinion via either direct-switch or cross-inhibition. Once the self-sourcing or polling phase is complete, the robot determines the exploration time and returns to the exploration state. A committed robot draws the exploration time from an exponential distribution with mean $\lambda^{-1}_e = t_e = 3\,000$, corresponding to an average exploration time of approximately 100 seconds. 
Throughout the paper, the values of $t_d$, $t_e$, and $t_u$ in the ODE models and the Gillespie simulations are fixed to those used in the robot simulations and experiments. All symbols and their definitions, together with the pseudocode describing the robot behaviour, are provided in the Supplementary Material (Table SL1 and Algorithm SA1).

\subsubsection{Real-robot demonstrations}
Following the same setup used in the simulations, we conducted a series of real-robot demonstrations using a swarm of 50 Kilobots in a \unit[$1 \times 1$]{m$^2$} square Kilogrid arena comprising 400 cells (Figure~\ref{fig:robot-fsm}E). Each trial was initialised with the swarm equally divided between the two opinions: 25 robots committed to the red option and 25 to the blue option. A demonstration was terminated when a quorum of $Q=0.7$ for either option was sustained for one minute, or after a maximum duration of 20 minutes.
Robots communicated their current opinion through their coloured LEDs (option $A$ as red, option $B$ as blue, and uncommitted as green). To limit the number of physical trials, we tested direct-switch and cross-inhibition under asocial dynamics type-2 (the more challenging variant, in which robots directly adopt the asocially sourced opinion rather than becoming uncommitted). We conducted selected experiments with three repetitions for each condition of $\eta_a \in \{0.3,0.5\}$ for $q\in \{0.66,0.92\}$, for a total of 24 experiments. The results, shown as black dots in Figures~\ref{fig:density_synergetic} and~\ref{fig:density_antagonistic}, are in good qualitative agreement with both the mathematical model and the robot simulations.


\subsection{Speed-accuracy and speed-cohesion metrics} \label{sec:metrics}

We analyse the relationship between decision speed, collective accuracy, and group cohesion using the following metrics.

\paragraph{Cohesion} reflects the degree to which the robots choose the same option. A run is said to be fully cohesive if all the robots (i.e., the whole swarm) choose the same option~\cite{franks|etal:2013}.
We quantify cohesion at termination time $T$ as the normalised absolute population difference between the number of robots committed to options A and B, that is, $\mathopen|$number robots for A -- number robots for B$\mathclose|$ / $N$, where $|.|$ is the absolute value operator. For each configuration of $q$, $\eta$, and $\eta_{a}$, we report the average cohesion across all runs.

\paragraph{Accuracy} represents the probability that the swarm selects the best option. For each tested configuration, we compute accuracy as the proportion of runs that reach quorum $Q=0.75$ for the best option A. A run is considered accurate when $A_D+A_E$ $\ge N\times Q$, i.e., when 75\% of the population has selected option A. When the two options have equal quality ($q=1$), reaching quorum $Q$ for either option is considered an accurate decision.

\paragraph{Decision speed} measures how quickly the swarm reaches a collective decision. We compute it as the mean number of time steps taken to reach quorum $Q=0.75$ for either option. Therefore, runs that fail to reach the quorum within the designated time limit $T$ are excluded from this calculation.

\section{Data Availability} 

The videos of the robot experiments are available at \url{https://iridia.ulb.ac.be/supp/IridiaSupp2026-001/}. The main video is also available at \url{https://youtu.be/CSirFypT9tY}.

\section{Code Availability} 
All simulation code, Mathematica notebooks, and robot control code are open source and publicly available in the GitHub repository: \url{https://github.com/rainazakir/asocialdynamics}.

\bibliography{cdm}

\begin{thebibliography}{10}
\urlstyle{rm}
\expandafter\ifx\csname url\endcsname\relax
  \def\url#1{\texttt{#1}}\fi
\expandafter\ifx\csname urlprefix\endcsname\relax\def\urlprefix{URL }\fi
\expandafter\ifx\csname doiprefix\endcsname\relax\def\doiprefix{DOI: }\fi
\providecommand{\bibinfo}[2]{#2}
\providecommand{\eprint}[2][]{\url{#2}}

\bibitem{book:Hamann:2018}
\bibinfo{author}{Hamann, H.}
\newblock \emph{\bibinfo{title}{Swarm Robotics: A Formal Approach}}
  (\bibinfo{publisher}{Springer Publishing Company, Incorporated},
  \bibinfo{year}{2018}), \bibinfo{edition}{1st} edn.

\bibitem{zapata2020}
\bibinfo{author}{Carrillo-Zapata, D.} \emph{et~al.}
\newblock \bibinfo{journal}{\bibinfo{title}{Mutual shaping in swarm robotics:
  User studies in fire and rescue, storage organization, and bridge
  inspection}}.
\newblock {\emph{\JournalTitle{Frontiers in Robotics and AI}}}
  \textbf{\bibinfo{volume}{7}}, \doiprefix\url{10.3389/frobt.2020.00053}
  (\bibinfo{year}{2020}).

\bibitem{hauert2014nano}
\bibinfo{author}{Hauert, S.} \& \bibinfo{author}{Bhatia, S.~N.}
\newblock \bibinfo{journal}{\bibinfo{title}{Mechanisms of cooperation in cancer
  nanomedicine: towards systems nanotechnology}}.
\newblock {\emph{\JournalTitle{Trends in Biotechnology}}}
  \textbf{\bibinfo{volume}{32}}, \bibinfo{pages}{448--455},
  \doiprefix\url{https://doi.org/10.1016/j.tibtech.2014.06.010}
  (\bibinfo{year}{2014}).
\newblock \bibinfo{note}{Special Issue: Next Generation Therapeutics}.

\bibitem{Tzoumas2024SI}
\bibinfo{author}{Tzoumas, G.}, \bibinfo{author}{Salina, L.},
  \bibinfo{author}{McConville, A.}, \bibinfo{author}{Richardson, T.} \&
  \bibinfo{author}{Hauert, S.}
\newblock \emph{\bibinfo{title}{Extinguishing Wildfires in Large Scale
  Scenarios Using Swarms of UAVs}}, \bibinfo{pages}{71--83}
  (\bibinfo{year}{2024}).

\bibitem{Valentini2017Review}
\bibinfo{author}{Valentini, G.}, \bibinfo{author}{Ferrante, E.} \&
  \bibinfo{author}{Dorigo, M.}
\newblock \bibinfo{journal}{\bibinfo{title}{The best-of-n problem in robot
  swarms: Formalization, state of the art, and novel perspectives}}.
\newblock {\emph{\JournalTitle{Frontiers in Robotics and AI}}}
  \textbf{\bibinfo{volume}{4}}, \bibinfo{pages}{9},
  \doiprefix\url{10.3389/frobt.2017.00009} (\bibinfo{year}{2017}).

\bibitem{seeley|buhrman:2001}
\bibinfo{author}{Seeley, T.~D.} \& \bibinfo{author}{Buhrman, S.~C.}
\newblock \bibinfo{journal}{\bibinfo{title}{Nest-site selection in honey bees:
  how well do swarms implement the "best-of-n" decision rule?}}
\newblock {\emph{\JournalTitle{Behavioral Ecology and Sociobiology}}}
  \textbf{\bibinfo{volume}{49}}, \bibinfo{pages}{416--427}
  (\bibinfo{year}{2001}).

\bibitem{usher2001}
\bibinfo{author}{Usher, M.} \& \bibinfo{author}{McClelland, J.~L.}
\newblock \bibinfo{journal}{\bibinfo{title}{The time course of perceptual
  choice: the leaky, competing accumulator model}}.
\newblock {\emph{\JournalTitle{Psychological review}}}
  \textbf{\bibinfo{volume}{108}}, \bibinfo{pages}{550--92},
  \doiprefix\url{10.1037//0033-295X.108.3.550} (\bibinfo{year}{2001}).

\bibitem{marshall2009}
\bibinfo{author}{Marshall, J. A.~R.} \emph{et~al.}
\newblock \bibinfo{journal}{\bibinfo{title}{On optimal decision-making in
  brains and social insect colonies}}.
\newblock {\emph{\JournalTitle{Journal of The Royal Society Interface}}}
  \textbf{\bibinfo{volume}{6}}, \bibinfo{pages}{1065--1074},
  \doiprefix\url{10.1098/rsif.2008.0511} (\bibinfo{year}{2009}).
\newblock
  \eprint{https://royalsocietypublishing.org/doi/pdf/10.1098/rsif.2008.0511}.

\bibitem{PIRRONE202266}
\bibinfo{author}{Pirrone, A.}, \bibinfo{author}{Reina, A.},
  \bibinfo{author}{Stafford, T.}, \bibinfo{author}{Marshall, J.~A.} \&
  \bibinfo{author}{Gobet, F.}
\newblock \bibinfo{journal}{\bibinfo{title}{Magnitude-sensitivity: rethinking
  decision-making}}.
\newblock {\emph{\JournalTitle{Trends in Cognitive Sciences}}}
  \textbf{\bibinfo{volume}{26}}, \bibinfo{pages}{66--80},
  \doiprefix\url{10.1016/j.tics.2021.10.006} (\bibinfo{year}{2022}).

\bibitem{Reina:scirep:2018}
\bibinfo{author}{Reina, A.}, \bibinfo{author}{Bose, T.},
  \bibinfo{author}{Trianni, V.} \& \bibinfo{author}{Marshall, J. A.~R.}
\newblock \bibinfo{journal}{\bibinfo{title}{Psychophysical laws and the
  superorganism}}.
\newblock {\emph{\JournalTitle{Scientific Reports}}}
  \textbf{\bibinfo{volume}{8}} (\bibinfo{year}{2018}).

\bibitem{Reina:SwInt:2021}
\bibinfo{author}{Reina, A.}, \bibinfo{author}{Ferrante, E.} \&
  \bibinfo{author}{Valentini, G.}
\newblock \bibinfo{journal}{\bibinfo{title}{Collective decision-making in
  living and artificial systems: editorial}}.
\newblock {\emph{\JournalTitle{Swarm Intelligence}}}
  \textbf{\bibinfo{volume}{15}}, \bibinfo{pages}{1--6},
  \doiprefix\url{10.1007/s11721-021-00195-5} (\bibinfo{year}{2021}).

\bibitem{march2024honeybee}
\bibinfo{author}{March-Pons, D.}, \bibinfo{author}{M{\'u}gica, J.},
  \bibinfo{author}{Ferrero, E.~E.} \& \bibinfo{author}{Miguel, M.~C.}
\newblock \bibinfo{journal}{\bibinfo{title}{Honeybee-like collective decision
  making in a kilobot swarm}}.
\newblock {\emph{\JournalTitle{Physical Review Research}}}
  \textbf{\bibinfo{volume}{6}}, \bibinfo{pages}{033149} (\bibinfo{year}{2024}).

\bibitem{Hunt2020}
\bibinfo{author}{Hunt, E.~R.} \& \bibinfo{author}{Hauert, S.}
\newblock \bibinfo{journal}{\bibinfo{title}{A checklist for safe robot
  swarms}}.
\newblock {\emph{\JournalTitle{Nature Machine Intelligence}}}
  \textbf{\bibinfo{volume}{2}}, \bibinfo{pages}{420--422},
  \doiprefix\url{10.1038/s42256-020-0213-2} (\bibinfo{year}{2020}).

\bibitem{parker|zhang:2009}
\bibinfo{author}{Parker, C. A.~C.} \& \bibinfo{author}{Zhang, H.}
\newblock \bibinfo{journal}{\bibinfo{title}{Cooperative decision-making in
  decentralized multiple-robot systems: The best-of-n problem}}.
\newblock {\emph{\JournalTitle{IEEE/ASME Transactions on Mechatronics}}}
  \textbf{\bibinfo{volume}{14}}, \bibinfo{pages}{240--251}
  (\bibinfo{year}{2009}).

\bibitem{marshall|bogacz|dornhaus|planque|kovacs|franks:2009}
\bibinfo{author}{Marshall, J. A.~R.} \emph{et~al.}
\newblock \bibinfo{journal}{\bibinfo{title}{On optimal decision-making in
  brains and social insect colonies}}.
\newblock {\emph{\JournalTitle{Journal of The Royal Society Interface}}}
  \textbf{\bibinfo{volume}{6}}, \bibinfo{pages}{1065--1074}
  (\bibinfo{year}{2009}).

\bibitem{valentini|hamann|dorigo:2014}
\bibinfo{author}{Valentini, G.}, \bibinfo{author}{Hamann, H.} \&
  \bibinfo{author}{Dorigo, M.}
\newblock \bibinfo{title}{Self-organized collective decision making: The
  weighted voter model}.
\newblock In \emph{\bibinfo{booktitle}{Proceedings of the 13th International
  Conference on Autonomous Agents and Multiagent Systems}}, AAMAS~'14,
  \bibinfo{pages}{45--52} (\bibinfo{publisher}{IFAAMAS},
  \bibinfo{address}{Richland, SC}, \bibinfo{year}{2014}).

\bibitem{valentini|ferrante|hamann|dorigo:2015}
\bibinfo{author}{Valentini, G.}, \bibinfo{author}{Ferrante, E.},
  \bibinfo{author}{Hamann, H.} \& \bibinfo{author}{Dorigo, M.}
\newblock \bibinfo{journal}{\bibinfo{title}{Collective decision with 100
  {K}ilobots: {S}peed versus accuracy in binary discrimination problems}}.
\newblock {\emph{\JournalTitle{Autonomous Agents and Multi-Agent Systems}}}
  \textbf{\bibinfo{volume}{30}}, \bibinfo{pages}{553--580}
  (\bibinfo{year}{2016}).

\bibitem{talamaliSciRob2021}
\bibinfo{author}{Talamali, M.~S.}, \bibinfo{author}{Saha, A.},
  \bibinfo{author}{Marshall, J. A.~R.} \& \bibinfo{author}{Reina, A.}
\newblock \bibinfo{journal}{\bibinfo{title}{When less is more: Robot swarms
  adapt better to changes with constrained communication}}.
\newblock {\emph{\JournalTitle{Science Robotics}}}
  \textbf{\bibinfo{volume}{6}}, \bibinfo{pages}{eabf1416},
  \doiprefix\url{10.1126/scirobotics.abf1416} (\bibinfo{year}{2021}).
\newblock \eprint{https://www.science.org/doi/pdf/10.1126/scirobotics.abf1416}.

\bibitem{shan2021discrete}
\bibinfo{author}{Shan, Q.} \& \bibinfo{author}{Mostaghim, S.}
\newblock \bibinfo{journal}{\bibinfo{title}{Discrete collective estimation in
  swarm robotics with distributed {B}ayesian belief sharing}}.
\newblock {\emph{\JournalTitle{Swarm Intelligence}}}
  \textbf{\bibinfo{volume}{15}}, \bibinfo{pages}{377--402}
  (\bibinfo{year}{2021}).

\bibitem{palina2019}
\bibinfo{author}{Bartashevich, P.} \& \bibinfo{author}{Mostaghim, S.}
\newblock \bibinfo{title}{Benchmarking collective perception: New task
  difficulty metrics for collective decision-making}.
\newblock In \bibinfo{editor}{Moura~Oliveira, P.}, \bibinfo{editor}{Novais, P.}
  \& \bibinfo{editor}{Reis, L.~P.} (eds.) \emph{\bibinfo{booktitle}{Progress in
  Artificial Intelligence}}, \bibinfo{pages}{699--711}
  (\bibinfo{publisher}{Springer International Publishing},
  \bibinfo{address}{Cham}, \bibinfo{year}{2019}).

\bibitem{pais|etal:2013}
\bibinfo{author}{Pais, D.} \emph{et~al.}
\newblock \bibinfo{journal}{\bibinfo{title}{A mechanism for value-sensitive
  decision-making}}.
\newblock {\emph{\JournalTitle{PLoS ONE}}} \textbf{\bibinfo{volume}{8}},
  \bibinfo{pages}{1--9}, \doiprefix\url{10.1371/journal.pone.0073216}
  (\bibinfo{year}{2013}).

\bibitem{reina|valentini|fernandezoto|dorigo|trianni:2015}
\bibinfo{author}{Reina, A.}, \bibinfo{author}{Valentini, G.},
  \bibinfo{author}{Fern\'andez-Oto, C.}, \bibinfo{author}{Dorigo, M.} \&
  \bibinfo{author}{Trianni, V.}
\newblock \bibinfo{journal}{\bibinfo{title}{A design pattern for decentralised
  decision making}}.
\newblock {\emph{\JournalTitle{PLoS ONE}}} \textbf{\bibinfo{volume}{10}},
  \bibinfo{pages}{e0140950}, \doiprefix\url{10.1371/journal.pone.0140950}
  (\bibinfo{year}{2015}).

\bibitem{Reina:PRE:2017}
\bibinfo{author}{Reina, A.}, \bibinfo{author}{Marshall, J. A.~R.},
  \bibinfo{author}{Trianni, V.} \& \bibinfo{author}{Bose, T.}
\newblock \bibinfo{journal}{\bibinfo{title}{Model of the best-of-{N} nest-site
  selection process in honeybees}}.
\newblock {\emph{\JournalTitle{Physical Review E}}}
  \textbf{\bibinfo{volume}{95}}, \bibinfo{pages}{052411},
  \doiprefix\url{10.1103/PhysRevE.95.052411} (\bibinfo{year}{2017}).

\bibitem{seeley|etal:2012}
\bibinfo{author}{Seeley, T.~D.} \emph{et~al.}
\newblock \bibinfo{journal}{\bibinfo{title}{Stop signals provide cross
  inhibition in collective decision-making by honeybee swarms}}.
\newblock {\emph{\JournalTitle{Science}}} \textbf{\bibinfo{volume}{335}},
  \bibinfo{pages}{108--111}, \doiprefix\url{10.1126/science.1210361}
  (\bibinfo{year}{2012}).

\bibitem{Cardelli2012}
\bibinfo{author}{Cardelli, L.} \& \bibinfo{author}{Csik{\'{a}}sz-Nagy, A.}
\newblock \bibinfo{journal}{\bibinfo{title}{The cell cycle switch computes
  approximate majority}}.
\newblock {\emph{\JournalTitle{Scientific Reports}}}
  \textbf{\bibinfo{volume}{2}}, \bibinfo{pages}{656},
  \doiprefix\url{10.1038/srep00656} (\bibinfo{year}{2012}).

\bibitem{Cardelli2017}
\bibinfo{author}{Cardelli, L.}, \bibinfo{author}{Hernansaiz-Ballesteros,
  R.~D.}, \bibinfo{author}{Dalchau, N.} \& \bibinfo{author}{Csik{\'{a}}sz-Nagy,
  A.}
\newblock \bibinfo{journal}{\bibinfo{title}{Efficient switches in biology and
  computer science}}.
\newblock {\emph{\JournalTitle{PLOS Computational Biology}}}
  \textbf{\bibinfo{volume}{13}}, \bibinfo{pages}{e1005100},
  \doiprefix\url{10.1371/journal.pcbi.1005100} (\bibinfo{year}{2017}).

\bibitem{Prasetyo2019}
\bibinfo{author}{Prasetyo, J.}, \bibinfo{author}{{De Masi}, G.} \&
  \bibinfo{author}{Ferrante, E.}
\newblock \bibinfo{journal}{\bibinfo{title}{{Collective decision making in
  dynamic environments}}}.
\newblock {\emph{\JournalTitle{Swarm Intelligence}}}
  \textbf{\bibinfo{volume}{13}}, \bibinfo{pages}{217--243}
  (\bibinfo{year}{2019}).

\bibitem{zakir_miscommunication_nodate}
\bibinfo{author}{Zakir, R.}, \bibinfo{author}{Dorigo, M.} \&
  \bibinfo{author}{Reina, A.}
\newblock \bibinfo{title}{Miscommunication between robots can improve group
  accuracy in best-of-n decision-making}.
\newblock In \emph{\bibinfo{booktitle}{2024 IEEE/RSJ International Conference
  on Intelligent Robots and Systems, IROS'24}}, \bibinfo{pages}{9014--9021},
  \doiprefix\url{10.1109/iros58592.2024.10802464} (\bibinfo{publisher}{IEEE},
  \bibinfo{address}{Piscataway, NJ}, \bibinfo{year}{2024}).

\bibitem{Mobilia_2015}
\bibinfo{author}{Mobilia, M.}
\newblock \bibinfo{journal}{\bibinfo{title}{Nonlinear q-voter model with
  inflexible zealots}}.
\newblock {\emph{\JournalTitle{Physical Review E}}}
  \textbf{\bibinfo{volume}{92}}, \doiprefix\url{10.1103/physreve.92.012803}
  (\bibinfo{year}{2015}).

\bibitem{reina_cross-inhibition_2023}
\bibinfo{author}{Reina, A.}, \bibinfo{author}{Zakir, R.},
  \bibinfo{author}{De~Masi, G.} \& \bibinfo{author}{Ferrante, E.}
\newblock \bibinfo{journal}{\bibinfo{title}{Cross-inhibition leads to group
  consensus despite the presence of strongly opinionated minorities and asocial
  behaviour}}.
\newblock {\emph{\JournalTitle{Communications Physics}}}
  \textbf{\bibinfo{volume}{6}}, \bibinfo{pages}{236},
  \doiprefix\url{10.1038/s42005-023-01345-3} (\bibinfo{year}{2023}).

\bibitem{Khaluf2017noise}
\bibinfo{author}{Khaluf, Y.}, \bibinfo{author}{Pinciroli, C.},
  \bibinfo{author}{Valentini, G.} \& \bibinfo{author}{Hamann, H.}
\newblock \bibinfo{journal}{\bibinfo{title}{The impact of agent density on
  scalability in collective systems: noise-induced versus majority-based
  bistability}}.
\newblock {\emph{\JournalTitle{Swarm Intelligence}}}
  \textbf{\bibinfo{volume}{11}}, \bibinfo{pages}{155--179},
  \doiprefix\url{10.1007/s11721-017-0137-6} (\bibinfo{year}{2017}).

\bibitem{AntZakDorRei2024:aamas}
\bibinfo{author}{Antonic, N.}, \bibinfo{author}{Zakir, R.},
  \bibinfo{author}{Dorigo, M.} \& \bibinfo{author}{Reina, A.}
\newblock \bibinfo{title}{Collective robustness of heterogeneous
  decision-makers against stubborn individual}.
\newblock In \emph{\bibinfo{booktitle}{Proceedings of the 23rd International
  Conference on Autonomous Agents and Multiagent Systems (AAMAS 2024)}},
  \bibinfo{pages}{68--77}, \doiprefix\url{10.5555/3635637.3662853}
  (\bibinfo{publisher}{International Foundation for Autonomous Agents and
  Multiagent Systems}, \bibinfo{address}{Richland, SC}, \bibinfo{year}{2024}).

\bibitem{giulia2021}
\bibinfo{author}{De~Masi, G.} \emph{et~al.}
\newblock \bibinfo{journal}{\bibinfo{title}{Robot swarm democracy: the
  importance of informed individuals against zealots}}.
\newblock {\emph{\JournalTitle{Swarm Intelligence}}}
  \textbf{\bibinfo{volume}{15}}, \doiprefix\url{10.1007/s11721-021-00197-3}
  (\bibinfo{year}{2021}).

\bibitem{zakir2022}
\bibinfo{author}{Zakir, R.}, \bibinfo{author}{Dorigo, M.} \&
  \bibinfo{author}{Reina, A.}
\newblock \bibinfo{title}{Robot swarms break decision deadlocks in collective
  perception through cross-inhibition}.
\newblock In \bibinfo{editor}{Dorigo, M.} \emph{et~al.} (eds.)
  \emph{\bibinfo{booktitle}{Swarm Intelligence}}, \bibinfo{pages}{209--221}
  (\bibinfo{publisher}{Springer International Publishing},
  \bibinfo{address}{Cham}, \bibinfo{year}{2022}).

\bibitem{Golman2014}
\bibinfo{author}{Golman, R.}, \bibinfo{author}{Hagmann, D.} \&
  \bibinfo{author}{Miller, J.~H.}
\newblock \bibinfo{journal}{\bibinfo{title}{Polya's bees: A model of
  decentralized decision-making}}.
\newblock {\emph{\JournalTitle{Science Advances}}}
  \textbf{\bibinfo{volume}{1}}, \bibinfo{pages}{e1500253},
  \doiprefix\url{10.1126/sciadv.1500253} (\bibinfo{year}{2015}).

\bibitem{Jhawar2020}
\bibinfo{author}{Jhawar, J.} \emph{et~al.}
\newblock \bibinfo{journal}{\bibinfo{title}{Noise-induced schooling of fish}}.
\newblock {\emph{\JournalTitle{Nature Physics}}} \textbf{\bibinfo{volume}{16}},
  \bibinfo{pages}{488--493}, \doiprefix\url{10.1038/s41567-020-0787-y}
  (\bibinfo{year}{2020}).

\bibitem{Liu2025}
\bibinfo{author}{Liu, Z.}, \bibinfo{author}{Crosscombe, M.} \&
  \bibinfo{author}{Lawry, J.}
\newblock \bibinfo{journal}{\bibinfo{title}{Imprecise evidence in social
  learning}}.
\newblock {\emph{\JournalTitle{Swarm Intelligence}}}
  \textbf{\bibinfo{volume}{19}}, \bibinfo{pages}{1--27},
  \doiprefix\url{10.1007/s11721-024-00238-7} (\bibinfo{year}{2025}).

\bibitem{Canciani2019}
\bibinfo{author}{Canciani, F.}, \bibinfo{author}{Talamali, M.~S.},
  \bibinfo{author}{Marshall, J. A.~R.}, \bibinfo{author}{Bose, T.} \&
  \bibinfo{author}{Reina, A.}
\newblock \bibinfo{title}{{Keep calm and vote on: Swarm resiliency in
  collective decision making}}.
\newblock In \emph{\bibinfo{booktitle}{Proceedings of Workshop Resilient Robot
  Teams of the 2019 IEEE International Conference on Robotics and Automation
  (ICRA 2019)}}, \bibinfo{pages}{4} (\bibinfo{publisher}{IEEE Press},
  \bibinfo{address}{Piscataway, NJ}, \bibinfo{year}{2019}).

\bibitem{JuliaKlein2024}
\bibinfo{author}{Klein, J.}, \bibinfo{author}{d'Onofrio, A.} \&
  \bibinfo{author}{Petrov, T.}
\newblock \bibinfo{title}{Exploring consensus robustness in swarms
  with disruptive individuals}.
\newblock In \bibinfo{editor}{Margaria, T.} \& \bibinfo{editor}{Steffen, B.}
  (eds.) \emph{\bibinfo{booktitle}{Leveraging Applications of Formal Methods,
  Verification and Validation. Rigorous Engineering of Collective Adaptive
  Systems}}, \bibinfo{pages}{33--48} (\bibinfo{publisher}{Springer Nature
  Switzerland}, \bibinfo{address}{Cham}, \bibinfo{year}{2025}).

\bibitem{Khalil_2018}
\bibinfo{author}{Khalil, N.}, \bibinfo{author}{Miguel, M.~S.} \&
  \bibinfo{author}{Toral, R.}
\newblock \bibinfo{journal}{\bibinfo{title}{Zealots in the mean-field noisy
  voter model}}.
\newblock {\emph{\JournalTitle{Physical Review E}}}
  \textbf{\bibinfo{volume}{97}}, \doiprefix\url{10.1103/physreve.97.012310}
  (\bibinfo{year}{2018}).

\bibitem{damore2025}
\bibinfo{author}{d'Amore, F.} \& \bibinfo{author}{Ziccardi, I.}
\newblock \bibinfo{journal}{\bibinfo{title}{Phase transition of the 3-majority
  opinion dynamics with noisy interactions}}.
\newblock {\emph{\JournalTitle{Theoretical Computer Science}}}
  \textbf{\bibinfo{volume}{1028}}, \bibinfo{pages}{115030},
  \doiprefix\url{10.1016/j.tcs.2024.115030} (\bibinfo{year}{2025}).

\bibitem{franks2008AnimBeh}
\bibinfo{author}{Franks, N.~R.} \emph{et~al.}
\newblock \bibinfo{journal}{\bibinfo{title}{Can ant colonies choose a
  far-and-away better nest over an in-the-way poor one?}}
\newblock {\emph{\JournalTitle{Animal Behaviour}}}
  \textbf{\bibinfo{volume}{76}}, \bibinfo{pages}{323--334},
  \doiprefix\url{10.1016/j.anbehav.2008.02.009} (\bibinfo{year}{2008}).

\bibitem{leaf2024}
\bibinfo{author}{Leaf, J.} \& \bibinfo{author}{Adams, J.~A.}
\newblock \bibinfo{journal}{\bibinfo{title}{The effect of uneven and obstructed
  site layouts in best-of-n}}.
\newblock {\emph{\JournalTitle{Swarm Intelligence}}}
  \textbf{\bibinfo{volume}{18}}, \bibinfo{pages}{311--333},
  \doiprefix\url{10.1007/s11721-024-00236-9} (\bibinfo{year}{2024}).

\bibitem{Centola2018}
\bibinfo{author}{Centola, D.}, \bibinfo{author}{Becker, J.},
  \bibinfo{author}{Brackbill, D.} \& \bibinfo{author}{Baronchelli, A.}
\newblock \bibinfo{journal}{\bibinfo{title}{Experimental evidence for tipping
  points in social convention}}.
\newblock {\emph{\JournalTitle{Science}}} \textbf{\bibinfo{volume}{360}},
  \bibinfo{pages}{1116--1119}, \doiprefix\url{10.1126/science.aas8827}
  (\bibinfo{year}{2018}).

\bibitem{couzin2011}
\bibinfo{author}{Couzin, I.~D.} \emph{et~al.}
\newblock \bibinfo{journal}{\bibinfo{title}{Uninformed individuals promote
  democratic consensus in animal groups}}.
\newblock {\emph{\JournalTitle{Science}}} \textbf{\bibinfo{volume}{334}},
  \bibinfo{pages}{1578--1580} (\bibinfo{year}{2011}).

\bibitem{Rajendran2022}
\bibinfo{author}{Rajendran, H.}, \bibinfo{author}{Haluts, A.},
  \bibinfo{author}{Gov, N.~S.} \& \bibinfo{author}{Feinerman, O.}
\newblock \bibinfo{journal}{\bibinfo{title}{Ants resort to majority concession
  to reach democratic consensus in the presence of a persistent minority}}.
\newblock {\emph{\JournalTitle{Current Biology}}}
  \textbf{\bibinfo{volume}{32}}, \bibinfo{pages}{645--653.e8},
  \doiprefix\url{10.1016/j.cub.2021.12.013} (\bibinfo{year}{2022}).

\bibitem{Marvel2012}
\bibinfo{author}{Marvel, S.~A.}, \bibinfo{author}{Hong, H.},
  \bibinfo{author}{Papush, A.} \& \bibinfo{author}{Strogatz, S.~H.}
\newblock \bibinfo{journal}{\bibinfo{title}{Encouraging moderation: Clues from
  a simple model of ideological conflict}}.
\newblock {\emph{\JournalTitle{Physical Review Letters}}}
  \textbf{\bibinfo{volume}{109}}, \bibinfo{pages}{118702},
  \doiprefix\url{10.1103/PhysRevLett.109.118702} (\bibinfo{year}{2012}).

\bibitem{gillespie2013perspective}
\bibinfo{author}{Gillespie, D.~T.}, \bibinfo{author}{Hellander, A.} \&
  \bibinfo{author}{Petzold, L.~R.}
\newblock \bibinfo{journal}{\bibinfo{title}{Perspective: Stochastic algorithms
  for chemical kinetics}}.
\newblock {\emph{\JournalTitle{The Journal of chemical physics}}}
  \textbf{\bibinfo{volume}{138}} (\bibinfo{year}{2013}).

\bibitem{Hamann:ANTS:2026}
\bibinfo{author}{Hamann, H.} \& \bibinfo{author}{Reina, A.}
\newblock \bibinfo{title}{Ten years of the collective perception benchmark in
  swarm robotics: Achievements and challenges}.
\newblock In \bibinfo{editor}{{R. Gross et al.}} (ed.)
  \emph{\bibinfo{booktitle}{Swarm Intelligence (ANTS 2026)}}, vol.
  \bibinfo{volume}{16515} of \emph{\bibinfo{series}{LNCS}}, \bibinfo{pages}{in
  press} (\bibinfo{publisher}{Springer}, \bibinfo{address}{Cham},
  \bibinfo{year}{2026}).

\bibitem{chinIROS2023ImperfectCPerception}
\bibinfo{author}{Chin, K.~Y.}, \bibinfo{author}{Khaluf, Y.} \&
  \bibinfo{author}{Pinciroli, C.}
\newblock \bibinfo{title}{Minimalistic collective perception with imperfect
  sensors}.
\newblock In \emph{\bibinfo{booktitle}{2023 IEEE/RSJ International Conference
  on Intelligent Robots and Systems (IROS)}}, \bibinfo{pages}{8862--8868},
  \doiprefix\url{10.1109/IROS55552.2023.10341384} (\bibinfo{year}{2023}).

\bibitem{valentini|brambilla|hamann|dorigo:2016}
\bibinfo{author}{Valentini, G.}, \bibinfo{author}{Brambilla, D.},
  \bibinfo{author}{Hamann, H.} \& \bibinfo{author}{Dorigo, M.}
\newblock \bibinfo{title}{Collective perception of environmental features in a
  robot swarm}.
\newblock In \bibinfo{editor}{Dorigo, M.} \& \bibinfo{editor}{{et al.}} (eds.)
  \emph{\bibinfo{booktitle}{Swarm Intelligence (ANTS 2016)}}, vol.
  \bibinfo{volume}{9882} of \emph{\bibinfo{series}{LNCS}},
  \bibinfo{pages}{65--76}, \doiprefix\url{10.1007/978-3-319-44427-7_6}
  (\bibinfo{publisher}{Springer}, \bibinfo{year}{2016}).

\bibitem{ebert2018multi}
\bibinfo{author}{Ebert, J.~T.}, \bibinfo{author}{Gauci, M.} \&
  \bibinfo{author}{Nagpal, R.}
\newblock \bibinfo{title}{Multi-feature collective decision making in robot
  swarms}.
\newblock In \emph{\bibinfo{booktitle}{Proceedings of the 17th International
  Conference on Autonomous Agents and MultiAgent Systems}},
  \bibinfo{pages}{1711--1719} (\bibinfo{year}{2018}).

\bibitem{rubenstein|ahler|hoff|cabrera|nagpal:2014}
\bibinfo{author}{Rubenstein, M.}, \bibinfo{author}{Ahler, C.},
  \bibinfo{author}{Hoff, N.}, \bibinfo{author}{Cabrera, A.} \&
  \bibinfo{author}{Nagpal, R.}
\newblock \bibinfo{journal}{\bibinfo{title}{Kilobot: A low cost robot with
  scalable operations designed for collective behaviors}}.
\newblock {\emph{\JournalTitle{Robotics and Autonomous Systems}}}
  \textbf{\bibinfo{volume}{62}}, \bibinfo{pages}{966--975}
  (\bibinfo{year}{2014}).

\bibitem{ValAntTra-etal2018:si}
\bibinfo{author}{Valentini, G.} \emph{et~al.}
\newblock \bibinfo{journal}{\bibinfo{title}{Kilogrid: A novel experimental
  environment for the kilobot robot}}.
\newblock {\emph{\JournalTitle{Swarm Intelligence}}}
  \textbf{\bibinfo{volume}{12}}, \bibinfo{pages}{245--266}
  (\bibinfo{year}{2018}).

\bibitem{vankampen:1992}
\bibinfo{author}{van Kampen, N.~G.}
\newblock \emph{\bibinfo{title}{Stochastic processes in physics and chemistry}}
  (\bibinfo{publisher}{Elsevier}, \bibinfo{address}{Amsterdam, NL},
  \bibinfo{year}{1992}).

\bibitem{franks|dornhaus|fitzsimmons|stevens:2003}
\bibinfo{author}{Franks, N.~R.}, \bibinfo{author}{Dornhaus, A.},
  \bibinfo{author}{Fitzsimmons, J.~P.} \& \bibinfo{author}{Stevens, M.}
\newblock \bibinfo{journal}{\bibinfo{title}{Speed versus accuracy in collective
  decision making}}.
\newblock {\emph{\JournalTitle{Proceedings of the Royal Society B: Biological
  Sciences}}} \textbf{\bibinfo{volume}{270}}, \bibinfo{pages}{2457--2463}
  (\bibinfo{year}{2003}).

\bibitem{CouzinEtAl2005}
\bibinfo{author}{Couzin, I.}, \bibinfo{author}{Krause, J.},
  \bibinfo{author}{Franks, N.} \& \bibinfo{author}{Levin, S.}
\newblock \bibinfo{journal}{\bibinfo{title}{Effective leadership and decision
  making in animal groups on the move}}.
\newblock {\emph{\JournalTitle{Nature}}} \textbf{\bibinfo{volume}{433}},
  \bibinfo{pages}{513--516} (\bibinfo{year}{2005}).

\bibitem{talamali2019improving}
\bibinfo{author}{Talamali, M.~S.}, \bibinfo{author}{Marshall, J.~A.},
  \bibinfo{author}{Bose, T.} \& \bibinfo{author}{Reina, A.}
\newblock \bibinfo{title}{Improving collective decision accuracy via
  time-varying cross-inhibition}.
\newblock In \emph{\bibinfo{booktitle}{2019 International conference on
  robotics and automation (ICRA)}}, \bibinfo{pages}{9652--9659}
  (\bibinfo{organization}{IEEE}, \bibinfo{year}{2019}).

\bibitem{franks|etal:2013}
\bibinfo{author}{Franks, N.~R.} \emph{et~al.}
\newblock \bibinfo{journal}{\bibinfo{title}{Speed--cohesion trade-offs in
  collective decision making in ants and the concept of precision in animal
  behaviour}}.
\newblock {\emph{\JournalTitle{Animal Behaviour}}}
  \textbf{\bibinfo{volume}{85}}, \bibinfo{pages}{1233--1244}
  (\bibinfo{year}{2013}).

\bibitem{Reina:PRE:2024}
\bibinfo{author}{Reina, A.}, \bibinfo{author}{Njougouo, T.},
  \bibinfo{author}{Tuci, E.} \& \bibinfo{author}{Carletti, T.}
\newblock \bibinfo{journal}{\bibinfo{title}{Speed-accuracy trade-offs in
  best-of-n collective decision making through heterogeneous mean-field
  modeling}}.
\newblock {\emph{\JournalTitle{Physical Review E}}}
  \textbf{\bibinfo{volume}{109}}, \bibinfo{pages}{054307},
  \doiprefix\url{10.1103/PhysRevE.109.054307} (\bibinfo{year}{2024}).

\bibitem{king2007bioletters}
\bibinfo{author}{King, A.~J.} \& \bibinfo{author}{Cowlishaw, G.}
\newblock \bibinfo{journal}{\bibinfo{title}{When to use social information: the
  advantage of large group size in individual decision making}}.
\newblock {\emph{\JournalTitle{Biology Letters}}} \textbf{\bibinfo{volume}{3}},
  \bibinfo{pages}{137--139}, \doiprefix\url{10.1098/rsbl.2007.0017}
  (\bibinfo{year}{2007}).

\bibitem{hamann2022scalability}
\bibinfo{author}{Hamann, H.} \& \bibinfo{author}{Reina, A.}
\newblock \bibinfo{journal}{\bibinfo{title}{Scalability in computing and
  robotics}}.
\newblock {\emph{\JournalTitle{IEEE Transactions on Computers}}}
  \textbf{\bibinfo{volume}{71}}, \bibinfo{pages}{1453--1465},
  \doiprefix\url{10.1109/TC.2021.3089044} (\bibinfo{year}{2022}).

\bibitem{salahshourphyrev2019}
\bibinfo{author}{Salahshour, M.}
\newblock \bibinfo{journal}{\bibinfo{title}{Phase diagram and optimal
  information use in a collective sensing system}}.
\newblock {\emph{\JournalTitle{Phys. Rev. Lett.}}}
  \textbf{\bibinfo{volume}{123}}, \bibinfo{pages}{068101},
  \doiprefix\url{10.1103/PhysRevLett.123.068101} (\bibinfo{year}{2019}).

\bibitem{coleman2021}
\bibinfo{author}{Bak-Coleman, J.~B.} \emph{et~al.}
\newblock \bibinfo{journal}{\bibinfo{title}{Stewardship of global collective
  behavior}}.
\newblock {\emph{\JournalTitle{Proceedings of the National Academy of
  Sciences}}} \textbf{\bibinfo{volume}{118}}, \bibinfo{pages}{e2025764118},
  \doiprefix\url{10.1073/pnas.2025764118} (\bibinfo{year}{2021}).

\bibitem{Borofsky2020}
\bibinfo{author}{Borofsky, T.} \emph{et~al.}
\newblock \bibinfo{journal}{\bibinfo{title}{Hive minded: like neurons, honey
  bees collectively integrate negative feedback to regulate decisions}}.
\newblock {\emph{\JournalTitle{Animal Behaviour}}}
  \textbf{\bibinfo{volume}{168}}, \bibinfo{pages}{33--44},
  \doiprefix\url{10.1016/j.anbehav.2020.07.023} (\bibinfo{year}{2020}).

\bibitem{Sridhar2021}
\bibinfo{author}{Sridhar, V.~H.} \emph{et~al.}
\newblock \bibinfo{journal}{\bibinfo{title}{The geometry of decision-making in
  individuals and collectives}}.
\newblock {\emph{\JournalTitle{Proceedings of the National Academy of
  Sciences}}} \textbf{\bibinfo{volume}{118}}, \bibinfo{pages}{e2102157118},
  \doiprefix\url{10.1073/pnas.2102157118} (\bibinfo{year}{2021}).

\bibitem{OddiANTS2022}
\bibinfo{author}{Oddi, F.}, \bibinfo{author}{Cristofaro, A.} \&
  \bibinfo{author}{Trianni, V.}
\newblock \bibinfo{title}{Best-of-n collective decisions on a hierarchy}.
\newblock In \emph{\bibinfo{booktitle}{Swarm Intelligence (ANTS 2022)}}, vol.
  \bibinfo{volume}{13491} of \emph{\bibinfo{series}{LNCS}},
  \bibinfo{pages}{66--78} (\bibinfo{publisher}{Springer},
  \bibinfo{address}{Cham}, \bibinfo{year}{2022}).

\bibitem{LeeANTS2018}
\bibinfo{author}{Lee, C.}, \bibinfo{author}{Lawry, J.} \&
  \bibinfo{author}{Winfield, A.}
\newblock \bibinfo{title}{Negative updating combined with opinion pooling in
  the best-of-n problem in swarm robotics}.
\newblock In \emph{\bibinfo{booktitle}{Swarm Intelligence}}, vol.
  \bibinfo{volume}{11172} of \emph{\bibinfo{series}{LNCS}},
  \bibinfo{pages}{97--108} (\bibinfo{publisher}{Springer},
  \bibinfo{address}{Cham}, \bibinfo{year}{2018}).

\bibitem{Reina:DARS:2016}
\bibinfo{author}{Reina, A.}, \bibinfo{author}{Bose, T.},
  \bibinfo{author}{Trianni, V.} \& \bibinfo{author}{Marshall, J. A.~R.}
\newblock \bibinfo{title}{Effects of spatiality on value-sensitive decisions
  made by robot swarms}.
\newblock In \emph{\bibinfo{booktitle}{Distributed Autonomous Robotic Systems
  (DARS 2016): The 13th International Symposium}}, vol.~\bibinfo{volume}{6} of
  \emph{\bibinfo{series}{{SPAR}}}, \bibinfo{pages}{461--473}
  (\bibinfo{publisher}{Springer International Publishing},
  \bibinfo{address}{Cham, Switzerland}, \bibinfo{year}{2018}).

\bibitem{franks|pratt|mallon|britton|sumpter:2002}
\bibinfo{author}{Franks, N.~R.}, \bibinfo{author}{Pratt, S.~C.},
  \bibinfo{author}{Mallon, E.~B.}, \bibinfo{author}{Britton, N.~F.} \&
  \bibinfo{author}{Sumpter, D. J.~T.}
\newblock \bibinfo{journal}{\bibinfo{title}{Information flow, opinion polling
  and collective intelligence in house-hunting social insects}}.
\newblock {\emph{\JournalTitle{Philosophical Transactions of the Royal Society
  B: Biological Sciences}}} \textbf{\bibinfo{volume}{357}},
  \bibinfo{pages}{1567--1583} (\bibinfo{year}{2002}).

\bibitem{Taleb2013}
\bibinfo{author}{Taleb, N.~N.}
\newblock \bibinfo{journal}{\bibinfo{title}{{'Antifragility' as a mathematical
  idea}}}.
\newblock {\emph{\JournalTitle{Nature}}} \textbf{\bibinfo{volume}{494}},
  \bibinfo{pages}{430--430}, \doiprefix\url{10.1038/494430e}
  (\bibinfo{year}{2013}).

\bibitem{leonard2024fast}
\bibinfo{author}{Leonard, N.~E.}, \bibinfo{author}{Bizyaeva, A.} \&
  \bibinfo{author}{Franci, A.}
\newblock \bibinfo{journal}{\bibinfo{title}{Fast and flexible multiagent
  decision-making}}.
\newblock {\emph{\JournalTitle{Annual Review of Control, Robotics, and
  Autonomous Systems}}} \textbf{\bibinfo{volume}{7}}.

\bibitem{gomez2023fish}
\bibinfo{author}{G{\'o}mez-Nava, L.} \emph{et~al.}
\newblock \bibinfo{journal}{\bibinfo{title}{Fish shoals resemble a stochastic
  excitable system driven by environmental perturbations}}.
\newblock {\emph{\JournalTitle{Nature Physics}}} \textbf{\bibinfo{volume}{19}},
  \bibinfo{pages}{663--669} (\bibinfo{year}{2023}).

\bibitem{dario2017}
\bibinfo{author}{Albani, D.}, \bibinfo{author}{IJsselmuiden, J.},
  \bibinfo{author}{Haken, R.} \& \bibinfo{author}{Trianni, V.}
\newblock \bibinfo{title}{Monitoring and mapping with robot swarms for
  agricultural applications}.
\newblock In \emph{\bibinfo{booktitle}{2017 14th IEEE International Conference
  on Advanced Video and Signal Based Surveillance (AVSS)}},
  \bibinfo{pages}{1--6}, \doiprefix\url{10.1109/AVSS.2017.8078478}
  (\bibinfo{year}{2017}).

\bibitem{Nagy2020}
\bibinfo{author}{Nagy, M.} \emph{et~al.}
\newblock \bibinfo{journal}{\bibinfo{title}{Synergistic benefits of group
  search in rats}}.
\newblock {\emph{\JournalTitle{Current Biology}}}
  \textbf{\bibinfo{volume}{30}}, \bibinfo{pages}{4733--4738.e4},
  \doiprefix\url{10.1016/j.cub.2020.08.079} (\bibinfo{year}{2020}).

\bibitem{Sosna2019}
\bibinfo{author}{Sosna, M. M.~G.} \emph{et~al.}
\newblock \bibinfo{journal}{\bibinfo{title}{Individual and collective encoding
  of risk in animal groups}}.
\newblock {\emph{\JournalTitle{Proceedings of the National Academy of
  Sciences}}} \textbf{\bibinfo{volume}{116}}, \bibinfo{pages}{20556--20561},
  \doiprefix\url{10.1073/pnas.1905585116} (\bibinfo{year}{2019}).

\bibitem{Rausch:SwInt:2019}
\bibinfo{author}{Rausch, I.}, \bibinfo{author}{Reina, A.},
  \bibinfo{author}{Simoens, P.} \& \bibinfo{author}{Khaluf, Y.}
\newblock \bibinfo{journal}{\bibinfo{title}{Coherent collective behaviour
  emerging from decentralised balancing of social feedback and noise}}.
\newblock {\emph{\JournalTitle{Swarm Intelligence}}}
  \textbf{\bibinfo{volume}{13}}, \bibinfo{pages}{321--345},
  \doiprefix\url{10.1007/s11721-019-00173-y} (\bibinfo{year}{2019}).

\bibitem{Becker2017}
\bibinfo{author}{Becker, J.}, \bibinfo{author}{Brackbill, D.} \&
  \bibinfo{author}{Centola, D.}
\newblock \bibinfo{journal}{\bibinfo{title}{Network dynamics of social
  influence in the wisdom of crowds}}.
\newblock {\emph{\JournalTitle{Proceedings of the National Academy of
  Sciences}}} \textbf{\bibinfo{volume}{114}}, \bibinfo{pages}{E5070--E5076},
  \doiprefix\url{10.1073/pnas.1615978114} (\bibinfo{year}{2017}).

\bibitem{strogatz2018}
\bibinfo{author}{Strogatz, S.~H.}
\newblock \emph{\bibinfo{title}{Nonlinear Dynamics and Chaos: With Applications
  to Physics, Biology, Chemistry, and Engineering}} (\bibinfo{publisher}{CRC
  Press}, \bibinfo{year}{2018}).

\bibitem{toral|tessone:2007}
\bibinfo{author}{Toral, R.} \& \bibinfo{author}{Tessone, C.~J.}
\newblock \bibinfo{journal}{\bibinfo{title}{Finite size effects in the dynamics
  of opinion formation}}.
\newblock {\emph{\JournalTitle{Communications in Computational Physics}}}
  \textbf{\bibinfo{volume}{2}}, \bibinfo{pages}{177--195}
  (\bibinfo{year}{2007}).

\bibitem{biancalani2014}
\bibinfo{author}{Biancalani, T.}, \bibinfo{author}{Dyson, L.} \&
  \bibinfo{author}{McKane, A.~J.}
\newblock \bibinfo{journal}{\bibinfo{title}{Noise-induced bistable states and
  their mean switching time in foraging colonies}}.
\newblock {\emph{\JournalTitle{Phys. Rev. Lett.}}}
  \textbf{\bibinfo{volume}{112}}, \bibinfo{pages}{038101},
  \doiprefix\url{10.1103/PhysRevLett.112.038101} (\bibinfo{year}{2014}).

\bibitem{Pinciroli2018}
\bibinfo{author}{Pinciroli, C.}, \bibinfo{author}{Talamali, M.~S.},
  \bibinfo{author}{Reina, A.}, \bibinfo{author}{Marshall, J. A.~R.} \&
  \bibinfo{author}{Trianni, V.}
\newblock \bibinfo{title}{Simulating {Kilobots} within {ARGoS}: models and
  experimental validation}.
\newblock In \bibinfo{editor}{{M. Dorigo et al.}} (ed.)
  \emph{\bibinfo{booktitle}{Swarm Intelligence (ANTS 2018)}}, vol.
  \bibinfo{volume}{11172} of \emph{\bibinfo{series}{LNCS}},
  \bibinfo{pages}{176--187},
  \doiprefix\url{https://doi.org/10.1007/978-3-030-00533-7_14}
  (\bibinfo{publisher}{Springer}, \bibinfo{address}{Cham},
  \bibinfo{year}{2018}).

\bibitem{TillANTS2022}
\bibinfo{author}{{Aust}, T.}, \bibinfo{author}{Talamali, M.},
  \bibinfo{author}{Dorigo, M.}, \bibinfo{author}{Hamann, H.} \&
  \bibinfo{author}{Reina, A.}
\newblock \bibinfo{title}{The hidden benefits of limited communication and slow
  sensing in collective monitoring of dynamic environments}.
\newblock In \bibinfo{editor}{Dorigo, M.} \& \bibinfo{editor}{{et al.}} (eds.)
  \emph{\bibinfo{booktitle}{Swarm Intelligence (ANTS 2022)}}, vol.
  \bibinfo{volume}{13491} of \emph{\bibinfo{series}{LNCS}}
  (\bibinfo{publisher}{Springer}, \bibinfo{year}{2022}).

\end{thebibliography}

\section*{Acknowledgements}

R.\,Zakir and M.\,Dorigo acknowledge support from the Belgian F.R.S.-FNRS, of which they are a FRIA Doctoral student and a Research Director, respectively. A.Reina acknowledges support from DFG under Germany's Excellence Strategy -- EXC 2117 - 422037984.

\section*{Author contributions statement}

A.R. conceived the original idea and directed the project. R.Z, T.C, and A.R. performed the mean-field ODE analysis. R.Z and A.R performed the stochastic analysis. R.Z. designed and implemented the robot control code for the robot simulations. R.Z. generated the figures. R.Z. and A.R wrote the first draft of the paper, and all authors edited the paper.

\end{document}